\documentclass{aa}  

\usepackage{amssymb}
\usepackage[usenames,dvipsnames]{xcolor}
\usepackage{graphicx}
\usepackage{nicefrac}
\usepackage[rightcaption]{sidecap}
\usepackage{natbib}
\usepackage{txfonts}
\usepackage[scaled=.86]{helvet}
\usepackage[normalem]{ulem}
\usepackage{booktabs}
\usepackage{url}
\usepackage{hyperref}
\hypersetup{
    final=true,
    pageanchor=true,
    colorlinks=true,
    breaklinks=true,
    linkcolor=blue,
    citecolor=blue,
    urlcolor=blue,   
    pdfpagemode=UseNone,
    pdftitle={GRIS/IFU pores},
    pdfauthor={M. Verma},
    pdfsubject={Solar Physics},
    pdfkeywords={Sun: activity -- photosphere -- chromosphere -- Line: profiles 
    -- Methods: observational}}
    
\usepackage{pifont}
\usepackage{epstopdf}

\definecolor{mvGreen}{cmyk}{0.992,0.,0.083, 0.525}

\begin{document} 
  
\title{High-resolution observations of two pores with the integral field 
    unit (IFU) of the GREGOR Infrared Spectrograph (GRIS)}

\author{%
    Meetu Verma}
    
\institute{%
    Leibniz-Institut f{\"u}r Astrophysik Potsdam (AIP),
    An der Sternwarte~16,
    14482 Potsdam,
    Germany,
%    \email{mverma@aip.de}
    \href{mailto:mverma@aip.de}{\textsf{mverma@aip.de}}}

\titlerunning{High-resolution observations of two pores}
\authorrunning{Verma}  
  
\date{Received 26 July 2023; accepted 19 June 2024}

\abstract
    {Solar pores are associated with two significant transitions in magnetohydrodynamics: the magnetic field becomes sufficiently strong to inhibit convective energy transport, and a critical change causes pores to develop a penumbra and to transform into sunspots.}
    {The goal is to compare the intricate details of the magnetic and flow fields around two solar pores, where one is part of an active region and the other is an isolated pore, with a secondary goal of demonstrating the scientific capabilities of the GRIS IFU.}
    {Two pores were observed with the High-resolution Fast Imager (HiFI) and the GRIS IFU at the 1.5-meter GREGOR solar telescope on 29~May and 6~June 2019. 
    %The first pore belongs to a bipolar region, whereas the other one is smaller and isolated. 
    The GRIS IFU mosaics provide spectropolarimetric data for inversions of the Ca\,\textsc{i} 1083.9~nm and Si\,\textsc{i} 1082.7~nm spectral lines, covering the deep and upper photosphere. The t-distributed Stochastic Neighbor Embedding (t-SNE) machine learning algorithm is employed to identify different classes of Si\,\textsc{i} Stokes-V profiles. The Local Correlation Tracking (LCT) technique derives horizontal proper motions around the pores using speckle-restored G-band time-series.}
    {Both pores contain a thin light bridge, are stable during the observations, and never develop a penumbra. The isolated pore is three times smaller and significantly darker than the active-region pore, which is not predicted by simulations. The LCT maps show inflows around both pores, with lower velocities for the isolated pore. Both pores are embedded in the photospheric LOS velocity pattern of the granulation but filamentary structures are only visible in the chromospheric LOS maps of the active-region pore. The t-SNE identifies five clusters of Si\,\textsc{i} Stokes-V profiles, revealing an `onion-peel' magnetic field structure, despite the small size of the pores. The core with strong vertical magnetic fields is surrounded by concentric layers with lower and more inclined magnetic fields. The GRIS IFU spectra allowed for the tracking of the temporal evolution of the physical parameters, but the variations for both pores were nominal.}
    {The active-region pore shows some signatures of increased interaction between plasma motions and magnetic fields, which can be considered as early signs of penumbra formation. However, similar physical properties prevail for smaller pores. Therefore, a statistically meaningful sample, covering the size range and different morphologies of pores, is needed to distinguish between the formation mechanisms of active-region and isolated pores.}

\keywords{Sun: activity -- evolution -- magnetic fields -- photosphere -- Line: profiles -- Methods: observational}
\maketitle

%===============================================================================
%    Introduction
%===============================================================================

\section{Introduction}

Magnetic fields occur on the solar surface at different spatial scales and in many different forms. They range from complex active regions to single magnetic flux tubes \citep{Steiner2001, Schuessler2003}. Solar pores are intermediate features between small-scale magnetic elements and sunspots. According to \citet{Rucklidge1995}, a certain overlap exists between large pores and small sunspots. The convective mode responsible for this overlap sets in suddenly and rapidly when the inclination to the vertical of the photospheric magnetic field exceeds some critical value. The consequence of this convective interchange is the filamentary penumbra, which governs the energy transport across the boundary of the sunspot into the external plasma. Pores are rarely found in quiet-Sun regions \citep{Verma2014}, and only about 10\% of pores exist in isolation. In general, pores do not have a circular shape. Typical aspect ratios of the semi-major and -minor axes are 3:2. Smaller pores tend to be more circular, and their boundaries are less corrugated \citep{Verma2014}.

The numerical modeling of \citet{Tildesley2004} provides confirmation of the presence of convectively driven filamentary instabilities, which play a role in the formation of sunspots. This process begins with small pores that accumulate magnetic flux through magnetic pumping. Theoretical models as well as observations show that the diameter of a pore increases with height in the atmosphere \citep[e.g.,][]{Simon1970, Zirin1992, Suetterlin1996, Keppens1996}. Various observations and models give varying values of the average field inclination in the range of 35\degr\,--\,70\degr\ \citep[e.g.,][]{Suetterlin1996, Keppens1996}.

Apart from the field inclination, the magnetic field strength is another important parameter for the transition from pore to sunspot. Analyzing data of the Solar Optical Telescope \citep[SOT,][]{Tsuneta2008} of the Japanese Hinode mission, \citet{Jurcak2018} suggested a fixed value of 1867~G at the umbra\,--\,penumbra boundary, providing a clear distinction between umbra and penumbra. \citet{Benko2018} found that this scenario does not hold for a sunspot with a decaying umbra. In a sunspot, where the penumbra decays, \citet{Verma2018b} found values lower than 1800~G for the vertical component of the magnetic field based on observations of the near-infrared Si\,\textsc{i} 1082.7~nm line. \citet{Lindner2020} carried out a statistical study to verify the existence of a constant value for the vertical component of the magnetic field at the umbra\,--\,penumbra boundary using data obtained with the GREGOR Infrared Spectrograph \citep[GRIS,][]{Collados2012}. This statistical study confirmed the existence of a constant value of about 1790~G, which slightly deviates from the one given by \citet{Jurcak2018}. However, this can be attributed to instrumental differences and the use of a different spectral line (Fe\,\textsc{i} 1564.8~nm). More recently, \citet{Loeptien2020} presented a statistical analysis of 23 sunspots observed with Hinode/SOT and concluded that no universal connection exists between the vertical magnetic field and the umbra\,--\,penumbra boundary in sunspots. They observed that the strength of the vertical magnetic field at the umbra\,--\,penumbra boundary depends predominantly on the size of the spots.

%With the advent of meter-class solar telescopes, the spatial, spectral, and temporal resolution of solar observations improved significantly.  

\citet{Keil1999} provided a qualitative characterization of velocities in the solar pores using a resolution of 0.5\arcsec\, and a sequence of about 40~min in different spectral lines and white light images. \citet{Sankarasubramanian2003} presented Doppler velocities using several spectral lines and magnetic field properties around the pores using slit scanning in the photospheric Fe\,\textsc{i} line pair at 630.2~nm. However, slit-reconstructed maps are not instantaneous so that features evolve while completing the scan. In addition, image restoration of solar spectra \citep{vanNoort2017} has not yet been widely adapted. \citet{Hirzberger2003} calculated the horizontal and Doppler velocities using imaging spectroscopy of non-magnetic lines. Image sequences from Hinode have been used, for example, by \citet{VargasDominguez2010} and \citet{Verma2014} to estimate horizontal proper motions around various pores. \citet{Sobotka2012} used imaging spectroploarimetric data in the Fe\,\textsc{i} 617.3~nm line to study magnetic and flow properties of a pore. \citet{Quintero2016} used the Hinode slit-scanner to study the photospheric magnetic field around a pore. They removed straylight using spatial deconvolution but the coarse data cadence hampered computing horizontal proper motions. Finally, \citet{Kamlah2023} used high-cadence H$\alpha$ imaging spectroscopy and photospheric images to study a solar pore just before it developed a penumbra. Despite different combinations of instruments, datasets, and spectral lines some innate properties of pores were found by most studies. However, none of these disparate datasets were able to provide a comprehensive description of the phenomenology and physics of solar pores. The current study aims to complement these studies, using newly available data of an integral field unit (IFU), which rendered high-cadence spectropolarimetric data of two pores in environments with different magnetic fields and plasma flows. The spectropolarimetric data were complemented by high-cadence imaging, which provides access to horizontal proper motions.

Using spectrographs with an IFU is common in observing the night sky. However, not many solar telescopes are equipped with such a device. The GRIS IFU \citep{VegaReyes2016, DominguezTagle2022} at the 1.5-meter GREGOR solar telescope \citep{Schmidt2012, Kleint2020} is one of the most recent examples. GRIS IFU data in the Fe\,\textsc{i} 1565~nm spectral range are already used to study quiet-Sun magnetic flux cancellation \citep{Kaithakkal2020}, the temporal evolution and characterization of the magnetic field vector related to the quiet-Sun internetwork \citep{Campbell2021}, and the oscillations of the magnetic field in pores \citep{Nelson2021}. \citet{Anan2021} presented the signatures of heating in plage regions using the He\,\textsc{i} triplet at 1083.0~nm and the Si\,\textsc{i} line at 1082.7~nm. However, so far no observations were presented in the spectral range of the He\,\textsc{i} triplet at 1083~nm covering strong magnetic features such as pores. Point of departure for this study is the finding by \citet{Verma2014} that only 10\% of all pores exist in isolation. They are formed by advection of flux by granular and supergranular motions. All other pores appear in bipolar or more complex active regions. Based on the working hypothesis that different mechanisms in different environments create these two types of pores, we will compare and, if possible, identify differences or similarities to further verify the hypothesis. This work is further motivated by new science capabilities offered by the GRIS IFU in the aforementioned spectral range, and the small size of pores makes them ideal targets for IFU spectropolarimetry. Observations and data reduction are described in Sect.~\ref{SEC02}. The temporal evolution of the magnetic field in the two pores will be discussed in Sect.~\ref{SEC03}. Section~\ref{SEC04} summarizes and evaluates the results and at the same time emphasizes their importance for future studies of pores.

%-------------------------------------------------------------------------------
%  HMI overview image
%-------------------------------------------------------------------------------
\begin{figure}[t]
\centering
\includegraphics[width=\columnwidth]{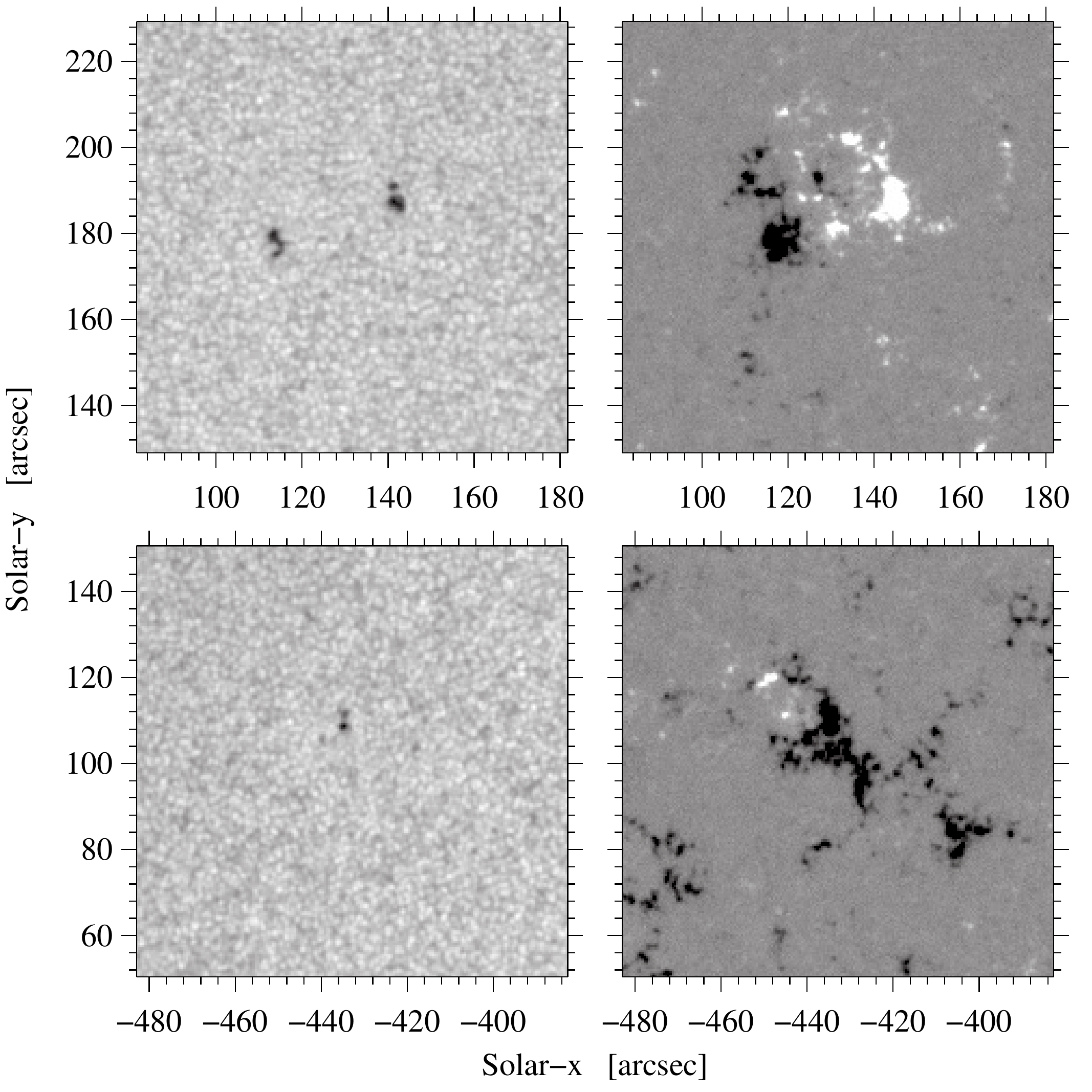}
\caption{Two pores analyzed in this study as observed in HMI continuum images (\textit{left}) and LOS magnetograms (\textit{right}) on 29~May 2019 (\textit{top}) and on 6~June 2019 (\textit{bottom}), respectively. The LOS magnetograms are displayed in the range of $\pm 200$~G.} 
\label{FIG01}
\end{figure}
%-------------------------------------------------------------------------------

%-------------------------------------------------------------------------------
%  HiFI images
%-------------------------------------------------------------------------------
\begin{figure*}[t]
\centering
\includegraphics[width=\textwidth]{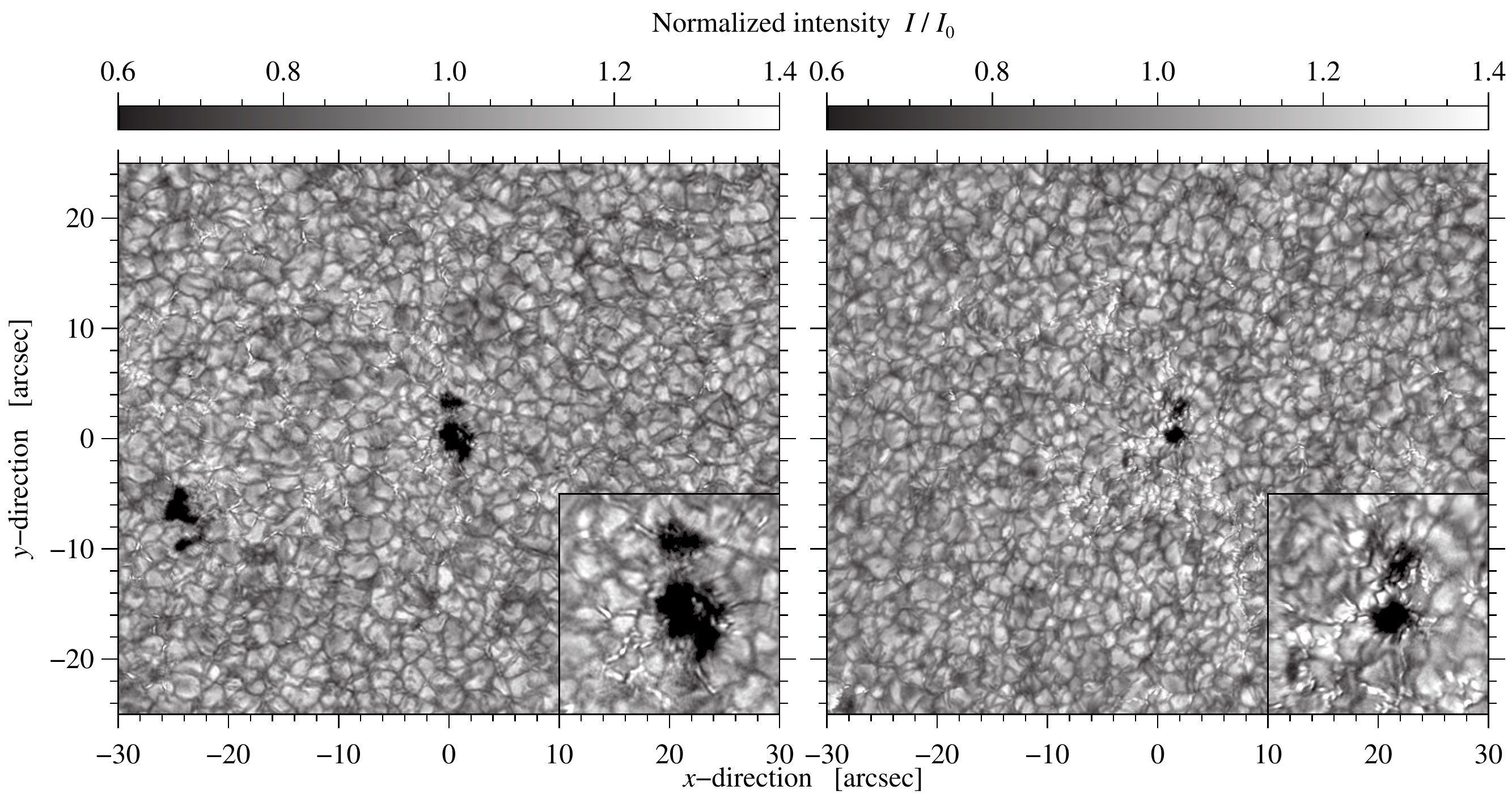}
\caption{Pores observed at 07:53~UT on 29~May (\textit{left}) and 
    at 07:42~UT on 6~June (\textit{right}) as seen in the best HiFI speckle-restored G-band images. A zoom-in with a FOV of 10\arcsec$\times$10\arcsec\, on the central pore is displayed in the lower right corners on both days. The images were normalized such that the mean of the quiet Sun intensity corresponds to unity. Animations showing the evolution of the pores are available online.} 
\label{FIG02}
\end{figure*}
%-------------------------------------------------------------------------------

%-------------------------------------------------------------------------------
%  LCT results
%-------------------------------------------------------------------------------
\begin{figure*}
\centering
\includegraphics[width=0.47\textwidth]{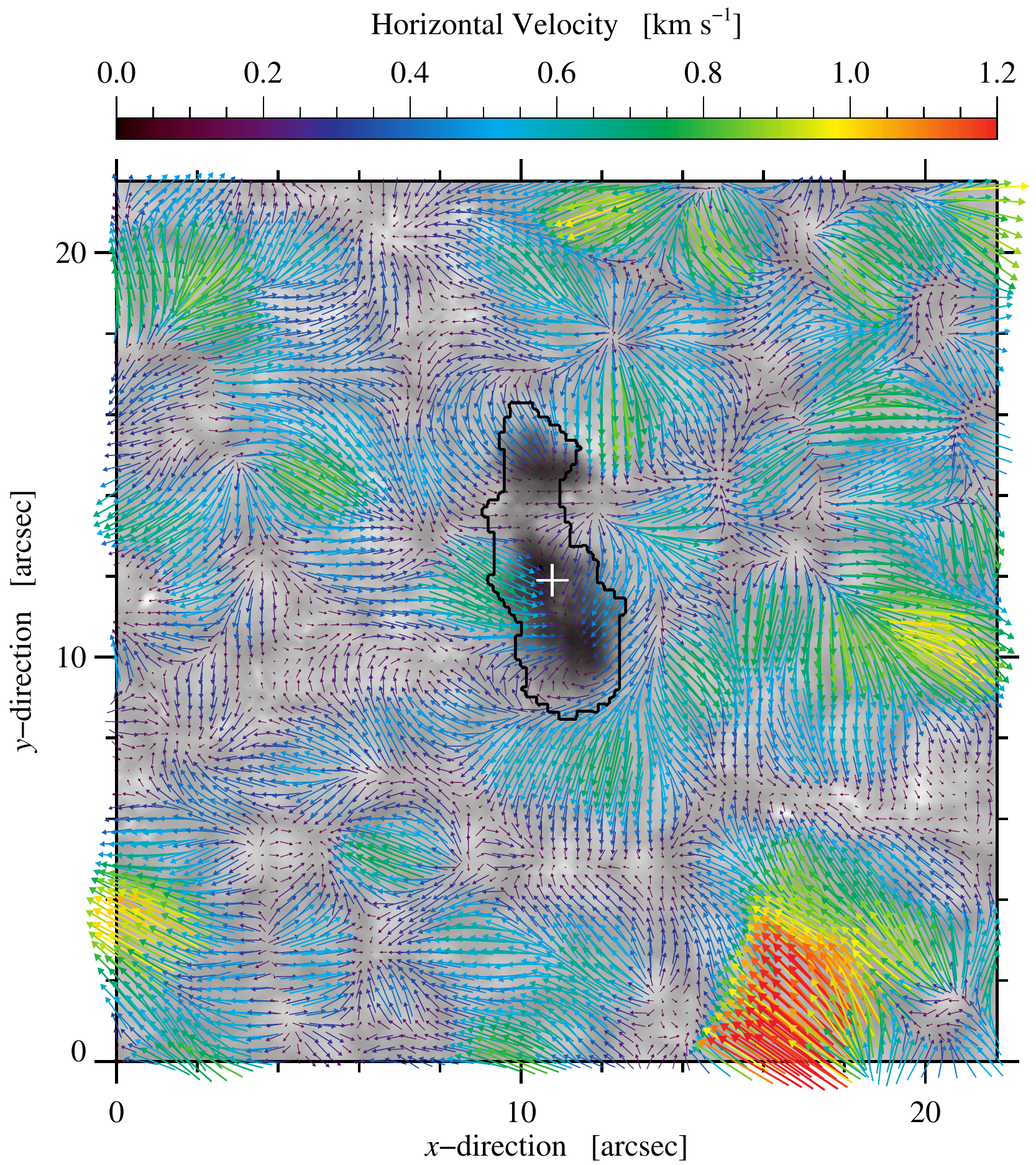}
\includegraphics[width=0.47\textwidth]{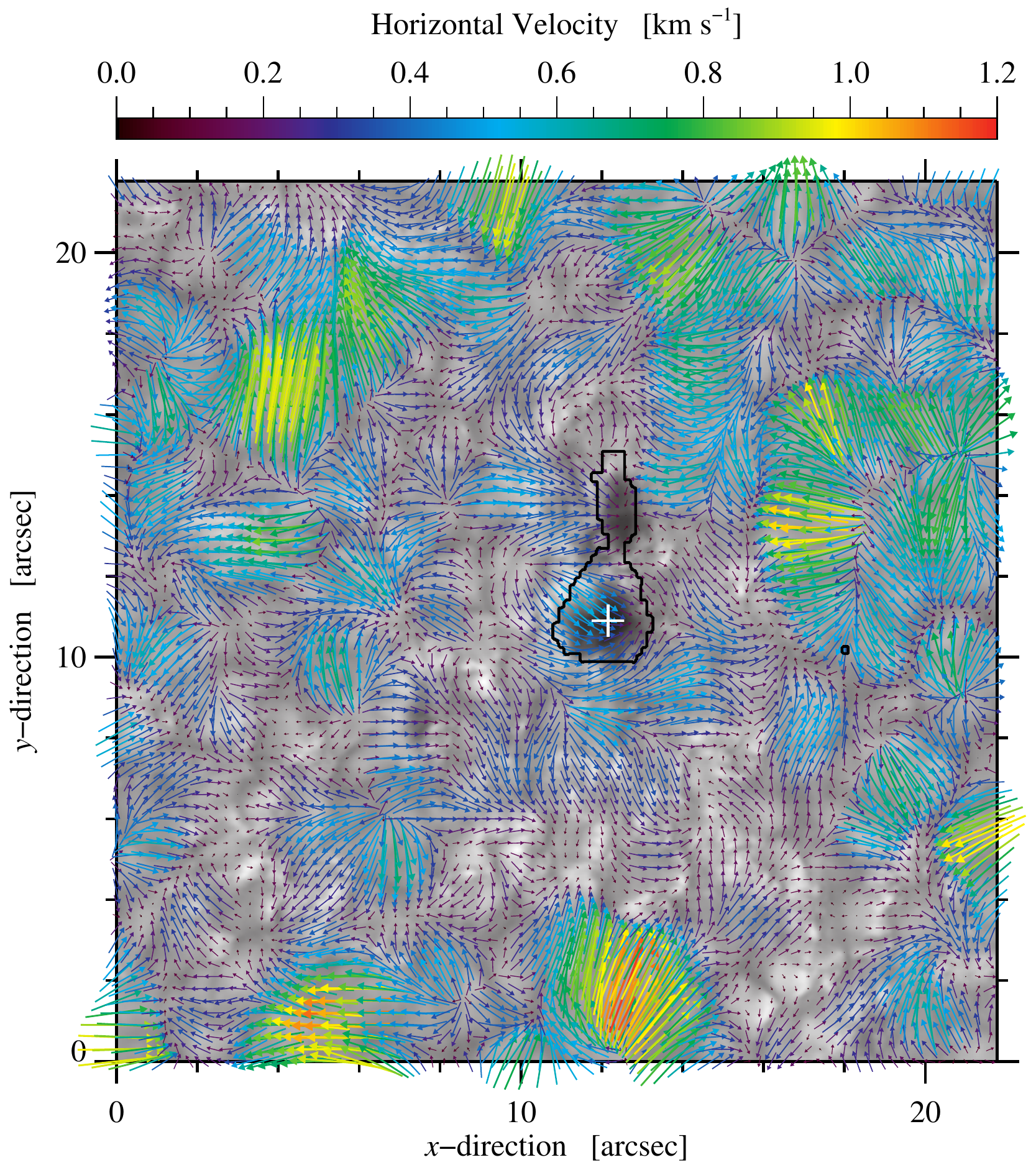}
\caption{Average horizontal proper motions based on LCT applied to HiFI G-band images acquired on 29~May (\textit{left}) and on 6~June (\textit{right}). The rainbow-colored arrows represent the direction and magnitude of the velocities. Black contours in the panel mark the pore borders. These contours are generated using GRIS continuum images. The white plus sign on both panels marks the centers of the pores around which the radial averages were taken.} 
\label{FIG14}
\end{figure*}
%-------------------------------------------------------------------------------

%-------------------------------------------------------------------------------
%  radial averages LCT velocities
%-------------------------------------------------------------------------------
\begin{figure*}
\centering
\includegraphics[width=0.47\textwidth]{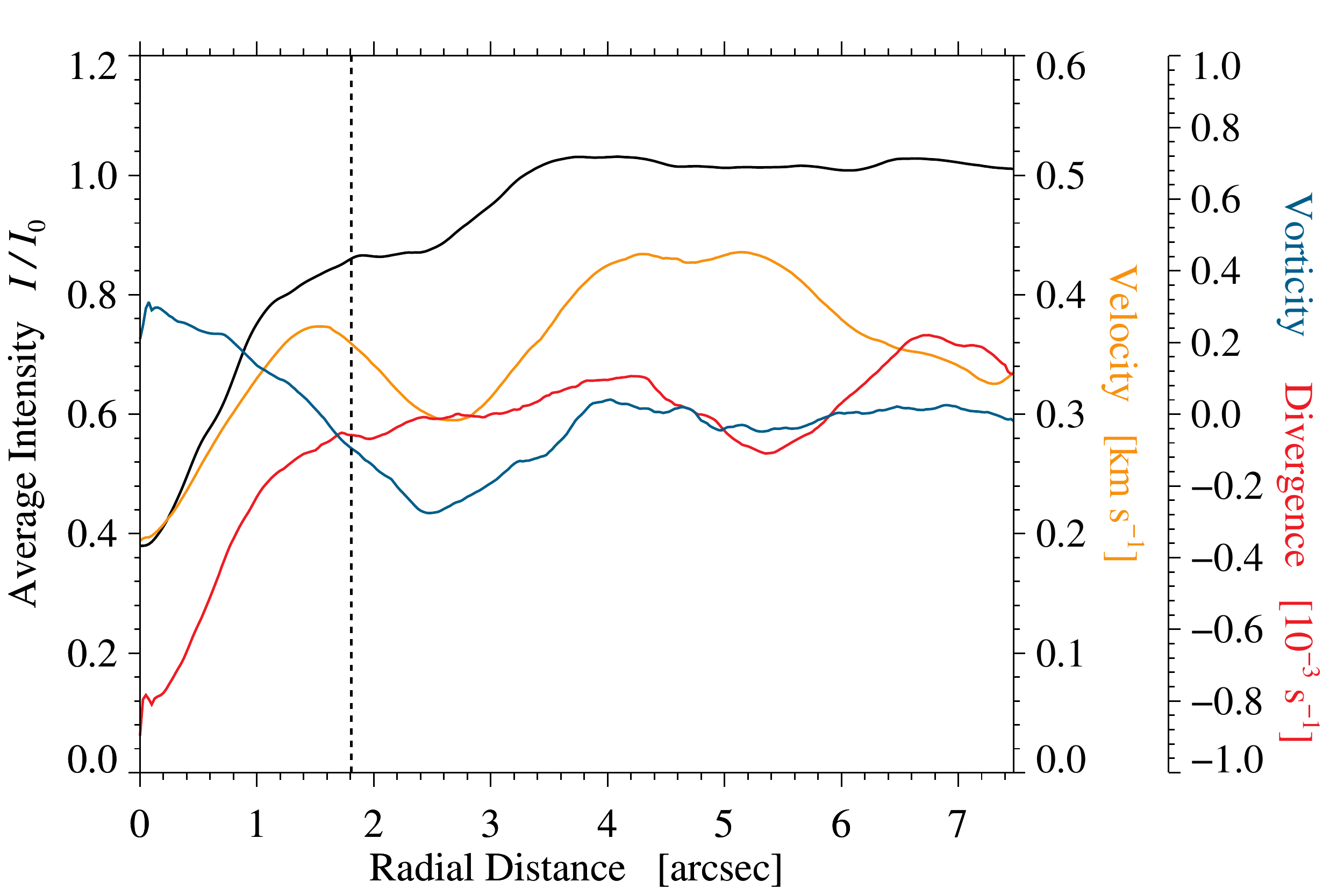}
\includegraphics[width=0.47\textwidth]{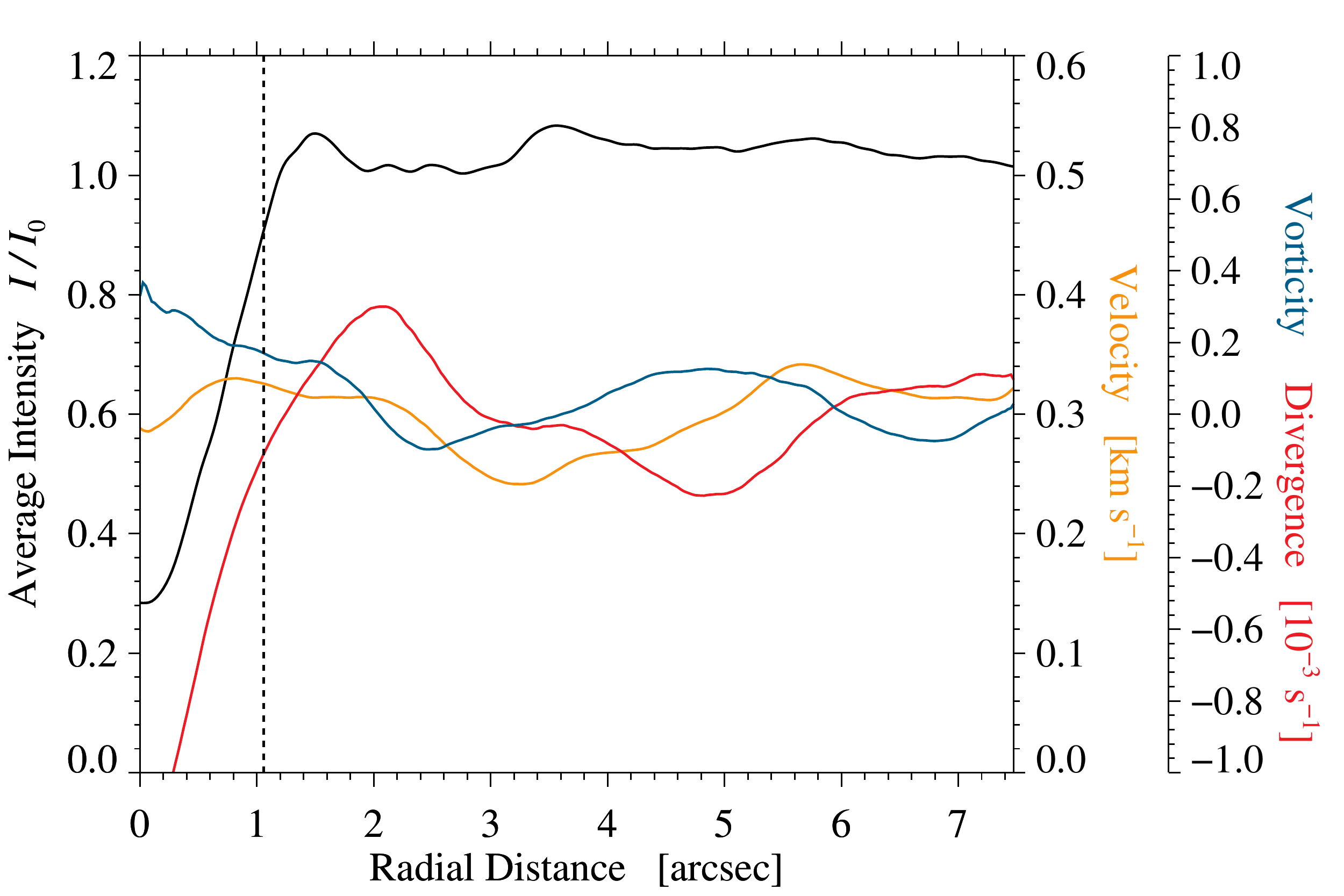}
\caption{Radial averages of normalized intensity (\textit{black}), horizontal
    velocity (\textit{orange}), divergence (\textit{red}), and vorticity (\textit{blue}) for pores observed on 29~May (\textit{left}) and on 6~June (\textit{right}). The center of the pores is at the origin. The equivalent pore radii (1.81\arcsec for 29~May and 1.06\arcsec for 6~June) are marked by vertical dashed-lines in both panels.} 
\label{FIG14a}
\end{figure*}
%-------------------------------------------------------------------------------

%===============================================================================
%    Observations
%===============================================================================

\section{Observations and data reduction\label{SEC02}}

% HMI - SDO
An overview of the two pores observed on 29~May and 6~June 2019 is provided based on continuum images and line-of-sight (LOS) magnetograms of the Helioseismic and Magnetic Imager \citep[HMI,][]{Scherrer2012} on board the Solar Dynamics Observatory \citep[SDO,][]{Pesnell2012}. The calibrated full-disk HMI continuum images and magnetograms are cropped to a field-of-view (FOV) of $100\arcsec \times 100\arcsec$ centered on the pores (Fig.~\ref{FIG01}). This FOV is sufficiently large to include the surrounding network in both cases.

% GRIS IFU
The spectropolarimetric data of the pores were obtained using the GRIS IFU. The FOV Scan System (FOVSS) is used for scanning across solar features. A single IFU exposure covers 6\arcsec $\times$ 3\arcsec. The detailed description of the IFU operation is given in \citet{DominguezTagle2022}. The wavelength range around 1083~nm covers the photospheric Si\,\textsc{i} spectral line at 1082.7~nm, the chromospheric He\,\textsc{i} triplet at 1083.0~nm, and the photospheric Ca\,\textsc{i} spectral line at 1083.9~nm. In this study, only two photospheric lines are used because the primary objective is to investigate the photospheric magnetic field of the two pores. The data on 29~May were taken as a $3 \times 3$ mosaic covering the leading pore in the emerging flux region located near disk center at $\mu=0.97$. The observations started at 07:52~UT and ended at 08:32~UT, resulting in a time-series of 20 mosaics covering a period of 40~min. The data on 6~June were taken as a $2 \times 2$ mosaic covering an isolated pore at $\mu=0.88$ in a decaying plage region. The observations started at 07:40~UT and ended at 08:07~UT, resulting in a time-series of 30 mosaics. The integration time used for each position in the mosaic was 100~ms with ten accumulations. The pixel scale is 0.135\arcsec\ pixel$^{-1}$ in the $x$-direction and 0.188\arcsec\ pixel$^{-1}$ in the $y$-direction. The assembled mosaic on 29~May covers $135 \times 48$~pixels corresponding to a FOV of $18.2\arcsec \times 9.0\arcsec$. The assembled mosaic on 6~June covers $90 \times 32$~pixels corresponding to a FOV of $12.2\arcsec \times 6.0\arcsec$. A spectral sampling of 1.80~pm pixel$^{-1}$ and 924 observed spectral points yield a spectral coverage of 1.66~nm on both observing days.

% HiFI rephrase
The two cameras of the High-resolution Fast Imager \citep[HiFI,][]{Denker2018a} recorded simultaneous high-spatial resolution images in the Fraunhofer G-band at 430.7~nm and in the blue continuum at 450.5~nm. The exposure time of 1.8~ms was the same for both cameras on both days. In general, sets of $2 \times 500$ frames were acquired with a frame rate of 47~Hz. After frame selection, only the best $2 \times 100$ calibrated frames were stored for post-processing and image restoration. The sCMOS imagers have a pixel size of 6.5~$\mu$m $\times$ 6.5~$\mu$m. The images have $2560 \times 2160$ pixels, and the pixel scale is about 0.025\arcsec\ pixel$^{-1}$, which leads to a FOV of about $64\arcsec \times 54\arcsec$. GREGOR HiFI data are processed using the \textit{sTools} software library \citep{Kuckein2017}. The speckle masking method is used for image restoration \citep{Woeger2008a, Woeger2008b}. 

The HiFI images on 29~May cover almost the same time period as the GRIS observations, i.e., sets of images were recorded at a cadence of 22~s from 07:53~UT to 08:33~UT. On 6~June, sets of images were also recorded at a cadence of 22~s, and the observed time period from 07:40~UT to 08:06~UT matches the GRIS data. Figure~\ref{FIG02} displays the best G-band images on 29~May and 6~June, respectively. The images are displayed with a FOV of $60\arcsec \times 50\arcsec$. The panels in the lower right corner zoom in on a region-of-interest (ROI) of $5\arcsec \times 5\arcsec$ containing the pores. The details seen in the HiFI blue continuum and G-band images are virtually identical. However, the bright points are more prominent in the G-band images. Therefore, only G-band images were used for further study. On both days the seeing was moderate with $r_0 = 8-10$~cm. Although the HiFI images are restored using the speckle masking technique, the blurring due to seeing is evident in the online animations. Some artifacts are present in multiple frames and occur when the restoration algorithm fails to correct frames with poor seeing. The GRIS IFU mosaics have not been restored or enhanced in any way. As a result, the seeing variations are quite evident and fine details are difficult to follow in the data. Note that HiFI and GRIS typically, record data simultaneously, except on 29~May, when the observer started the HiFI sequence about one minute later.

%-------------------------------------------------------------------------------
%    Data reduction and analysis
%-------------------------------------------------------------------------------

\begin{figure*}[t]
\centering
\includegraphics[width=\textwidth]{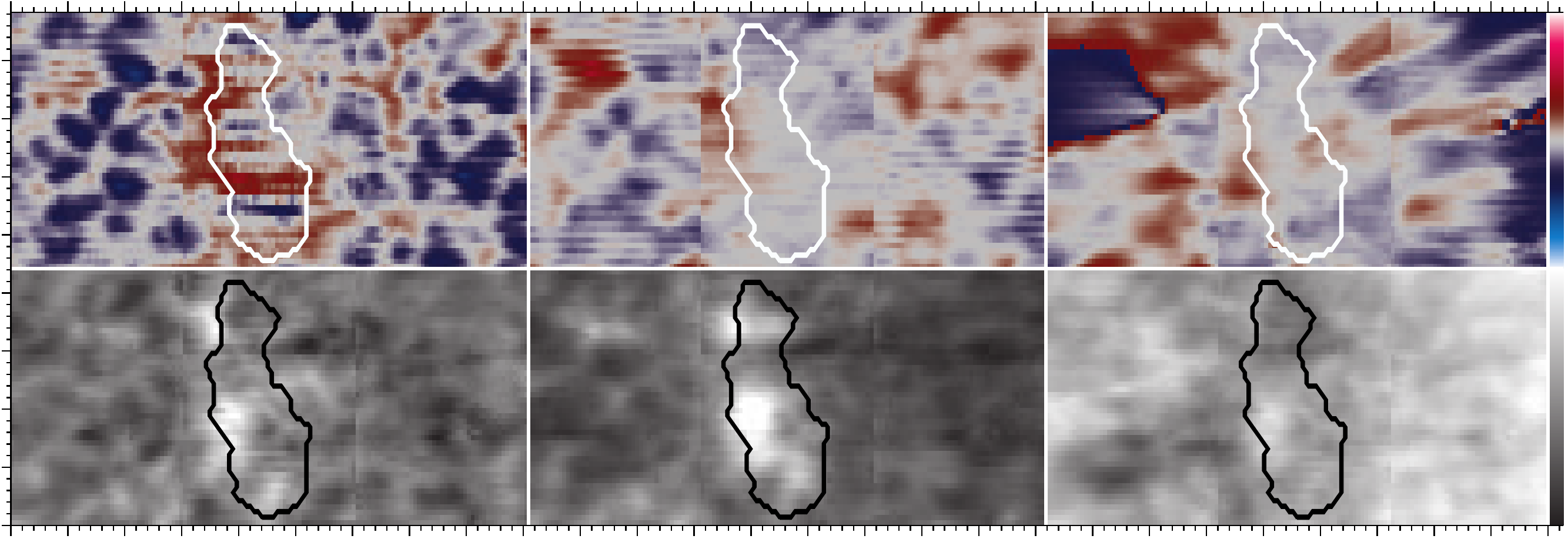}
\caption{LOS Doppler velocities (\textit{top}) and line-core intensity 
      (\textit{bottom}) derived from the GRIS IFU mosaic at
      the Ca\,\textsc{i}, Si\,\textsc{i}, and He\,\textsc{i} lines (\textit{left to right}) starting at 08:02~UT on 29~May. The major tick-marks represent 2\arcsec\ on both the $x$- and $y$-axis covering a FOV of 18\arcsec $\times$ 9\arcsec\, for each map. The color bar on the top-right shows the velocity range of $\pm$3~km~s$^{-1}$ for the Ca\,\textsc{i} and Si\,\textsc{i} lines and of $\pm$20~km~s$^{-1}$ for the He\,\textsc{i} lines. The color bar on the bottom-right shows the line-core intensity in the range of (0.80\,--\,1.04), (0.36\,--\,0.56), and (0.60\,--\,1.04) for the Ca\,\textsc{i}, Si\,\textsc{i}, and He\,\textsc{i} lines, respectively. The black and white contours outline the pore. The contours shown here and in all subsequent figures are extracted from the GRIS continuum mosaic at 08:02~UT on 29~May. The artifacts seen in the maps are discussed in Sect.~\ref{SEC02}. An animation showing the evolution of the pore is available online.} 
\label{FIG04}
\end{figure*}

\begin{figure*}[t]
\centering
\includegraphics[width=0.68\textwidth]{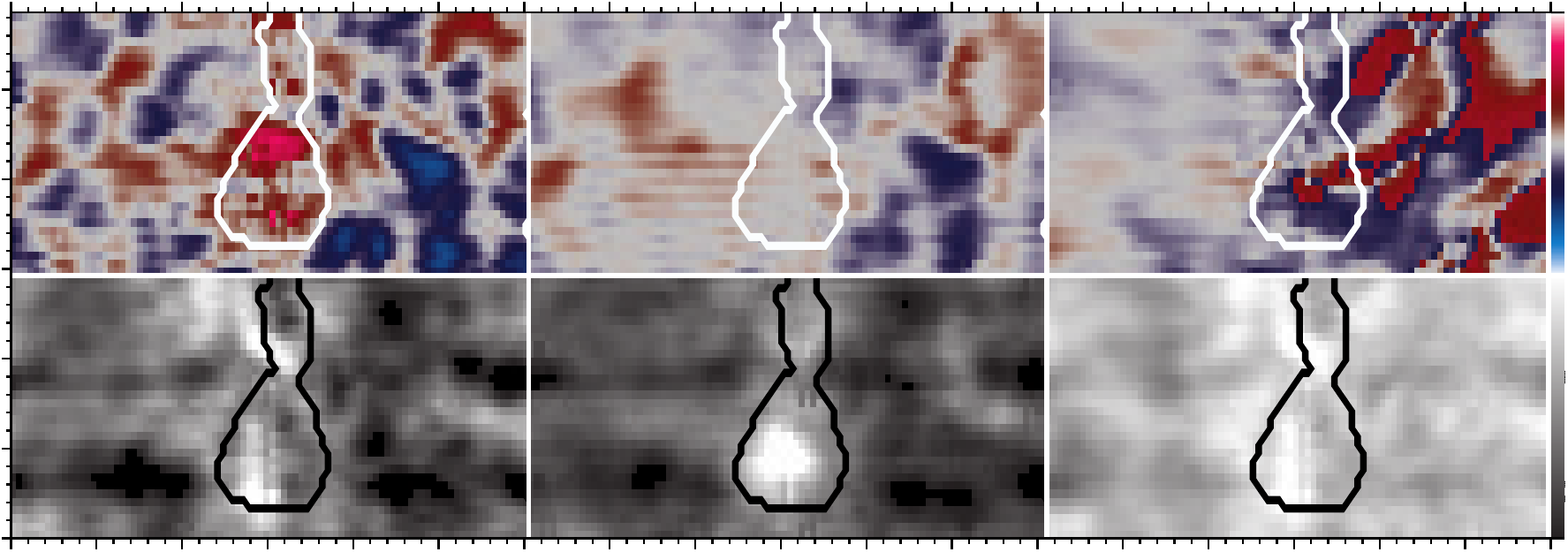}
\caption{LOS Doppler velocities (\textit{top}) and line-core intensity 
      (\textit{bottom}) derived from the GRIS IFU mosaic at
      the Ca\,\textsc{i}, Si\,\textsc{i}, and He\,\textsc{i} lines (\textit{left to right}) starting at 07:58~UT on 6~June. The major tick-marks represent 2\arcsec\ on both the $x$- and $y$-axis covering a FOV of 12\arcsec $\times$ 6\arcsec\, for each map. The color bar on the top-right shows the velocity range of $\pm$3~km~s$^{-1}$ for the Ca\,\textsc{i} and Si\,\textsc{i} lines and of $\pm$20~km~s$^{-1}$ for the He\,\textsc{i} lines. The color bar on the bottom-right shows the line-core intensity in the range of $(0.80-1.04)$, $(0.36-0.56)$, and $(0.60-1.04)$ for the Ca\,\textsc{i}, Si\,\textsc{i}, and He\,\textsc{i} lines, respectively. The black and white contours shown here and in all subsequent figures are extracted from the GRIS continuum mosaic at 07:58~UT on 6~June. The artifacts seen in the maps are discussed in Sect.~\ref{SEC02}. An animation showing the evolution of the pore is available online.} 
\label{FIG05}
\end{figure*}

% GRIS analysis and SIR inversions details
The data reduction for the GRIS IFU data is very similar to the one for GRIS slit observations. The GRIS pipeline runs on-site, once the daily observations are completed, and produces Level~1 data, which includes various processing steps such as dark-frame subtraction, flat-field correction, crosstalk removal, and calibration of the polarization modulator \citep{Collados1999}. In addition, the pipeline also assembles the spectral data into a 3D data cube that encompasses two spatial dimensions and the spectral dimension. The optical shifts introduced by the arrangement of mini-slits in the IFU are also removed by the pipeline \citep[see][]{DominguezTagle2022}. Further steps related to wavelength calibration and correction for the spectrograph profile, among others, are described in \citet{Verma2016b}. However, some artifacts are still present in the mosaic, as shown in Figs.~\ref{FIG04} and~\ref{FIG05}, which could be left over from the stitching of the IFU scan tiles. In addition, the FOVs covered by GRIS mosaics are very small. Neither smoothing was applied nor scalable pixels are used in the display, preserving the original spatial resolution. Thus, the images and related maps of physical properties appear pixelated. The inversion of the Si\,\textsc{i} line at 1082.7~nm and the Ca\,\textsc{i} line at 1083.9~nm ($g_\mathrm{eff}=1.5$ for both lines) is carried out with the ``Stokes Inversion based on Response functions" \citep[SIR,][]{RuizCobo1992} code. The computation of velocity and magnetic fields is implemented independent of height, hence the inversions of both lines are carried out separately \citep[cf.,][]{Verma2018b}. For the Si\,\textsc{i} and Ca\,\textsc{i} lines, the starting model covers the optical depth range $+1.0 \le \mathrm{log}\,\tau \le -5.4$ and $+1.0 \le \mathrm{log}\,\tau \le -4.4$, respectively. The limb-darkening factor is considered on both days according to Eq.~10 of \citet{Pierce1977b}, when estimating magnetic fields. A constant value of the macro-turbulence of 1~km~s$^{-1}$ and 2~km~s$^{-1}$  was assumed for the Si\,\textsc{i} and Ca\,\textsc{i} lines, respectively. Furthermore, a fixed stray light contribution of 2\% was used for both lines, which was chosen as the lower limit \citep[cf.,][]{Balthasar2016}. The inversions of both lines on both days yield a temperature stratification with three nodes $T(\tau)$ and the rest of the physical parameters including the total magnetic strength $B_\mathrm{tot}$, the magnetic inclination $\gamma$, the magnetic azimuth $\phi$, and the Doppler velocity $v_\mathrm{LOS}$ with only one node, i.e., constant with height.

% LCT
The horizontal proper motions are computed on both days using time-series of restored G-band images. Local correlation tracking \citep[LCT,][]{November1988, Verma2011, Verma2014} is employed with a Gaussian sampling window (FWHM $\approx 1200$~km), a time step of $\Delta t = 22$~s for computing individual flow maps, and a time period of $\Delta T = 39$~min and $28$~min on 29~May and on 6~June, respectively, for averaging the individual flow maps.

% t-SNE
To identify different classes for the available Stokes-V profiles, t-distributed Stochastic Neighbor Embedding \citep[t-SNE,][]{vanderMaaten2008} is used. The mathematical details about the method were presented by \citet{vanderMaaten2009, vanderMaaten2014}. The current set-up of t-SNE, i.e., the choice of the parameters (perplexity = 50, $\theta = 0.5$, and the number of iterations = 1000) is similar to the one described by \citet{Verma2021}, where t-SNE was used to classify H$\alpha$ profiles. The Stokes-V profiles of the Si\,\textsc{i} line served on both days as input data. In total 129\,600 and 86\,400 profiles are used on 29~May and on 6~June, respectively. Furthermore, clusters are recognized in the two-dimensional t-SNE projections with the Density-Based Spatial Clustering of Applications with Noise (DBSCAN) algorithm \citep{Ester1996}.

%-------------------------------------------------------------------------------
%  GRIS velocity
%-------------------------------------------------------------------------------
%-------------------------------------------------------------------------------

%-------------------------------------------------------------------------------
%  GRIS velocity
%-------------------------------------------------------------------------------

%-------------------------------------------------------------------------------

%-------------------------------------------------------------------------------
%  SIR inverted parameters for May 29
%-------------------------------------------------------------------------------
\begin{figure*}[t]
\centering
\includegraphics[width=\textwidth]{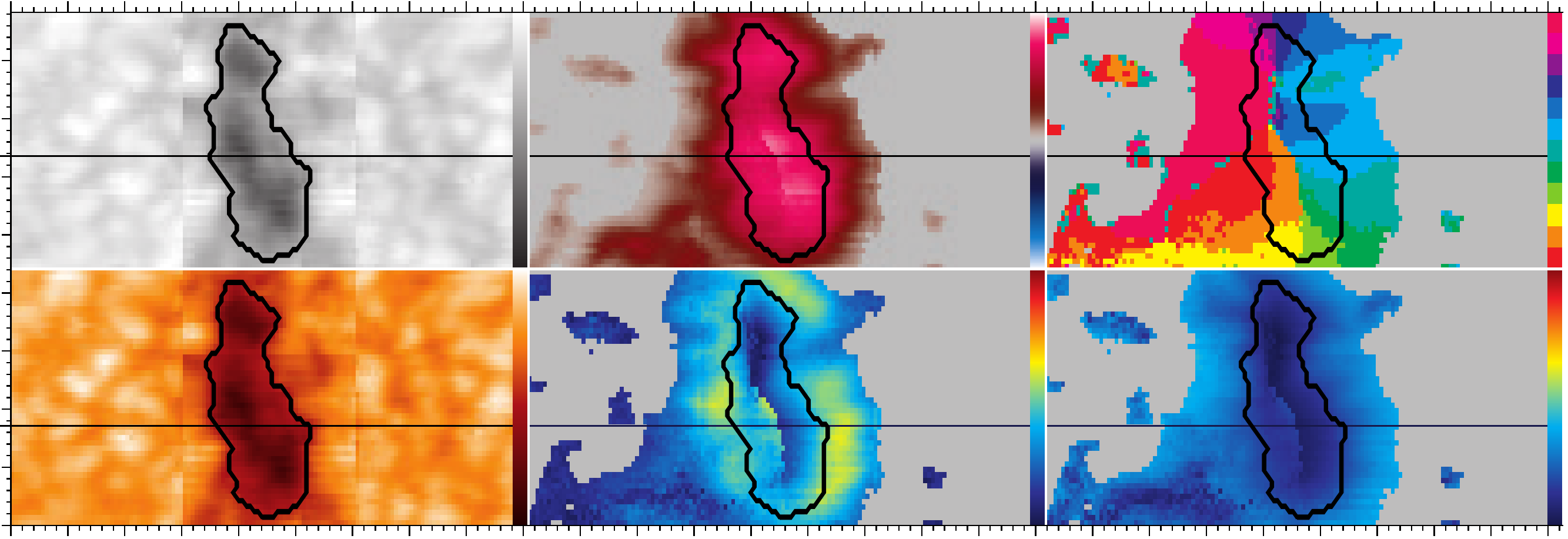}
\includegraphics[width=\textwidth]{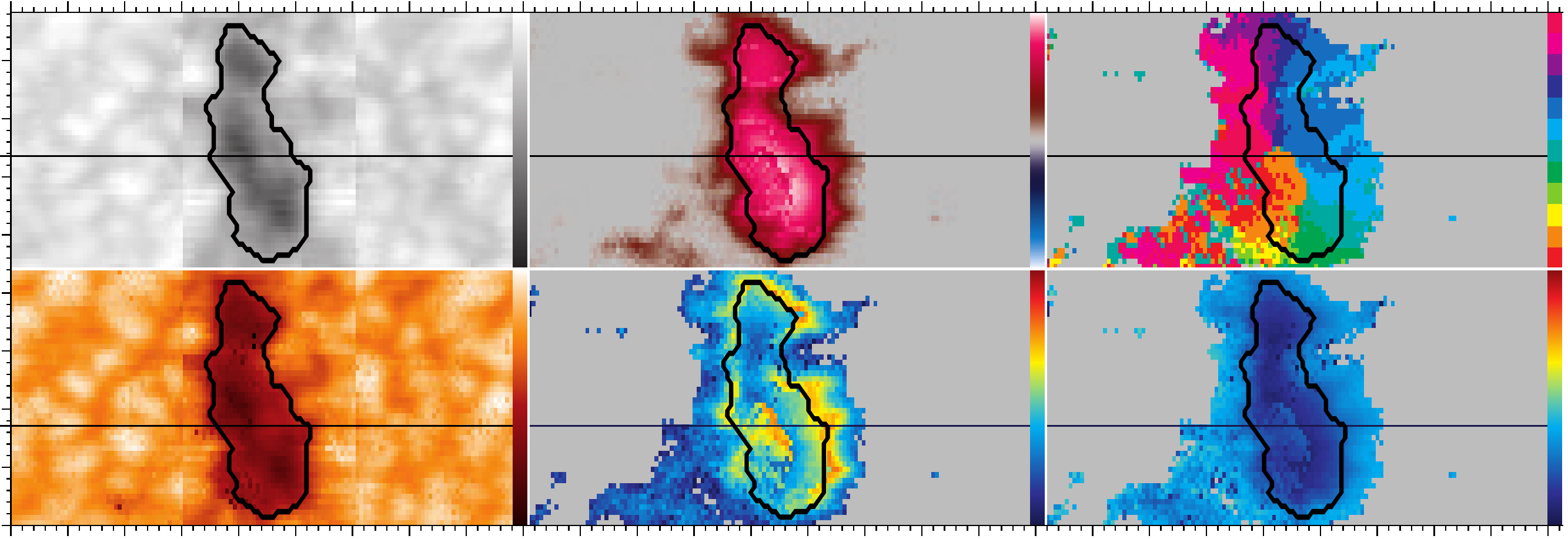}
\caption{Maps of six physical parameters derived for the Si\,\textsc{i} line 
    (\textit{top two rows}) and Ca\,\textsc{i} line (\textit{bottom two rows}) using the SIR code for the GRIS IFU mosaic starting at 08:02~UT on 29~May: normalized intensity $I / I_0$ ($0.7-1.05$), vertical component of magnetic field strength $B_z$ ($\pm$1500~G), magnetic field azimuth $\phi$ ($\pm$180\degr), temperature at $\tau = 0$ (5500~K\,--\,7000~K), horizontal component of magnetic field strength $B_\mathrm{hor}$ (0~G\,--\,1200~G), and magnetic field inclination $\gamma$ (0\degr\,--\,180\degr) (\textit{top-left to bottom-right}). The ranges for all physical parameters are depicted as colored bars on the right. The polarization signal in the gray region is below the noise level. Please note that $B_z$, $\phi$, and $\gamma$ are in the local reference frame. The major tick-marks represent 2\arcsec\ on both $x$- and $y$-axis covering the FOV of $18\arcsec \times 9\arcsec$ for each map. Animations showing the evolution of the pore are available online for the Si\,\textsc{i} and Ca\,\textsc{i} lines.} 
\label{FIG06}
\end{figure*}
%-------------------------------------------------------------------------------

%-------------------------------------------------------------------------------
%  SIR inverted parameters for June 06
%-------------------------------------------------------------------------------
\begin{figure*}[t]
\centering
\includegraphics[width=0.68\textwidth]{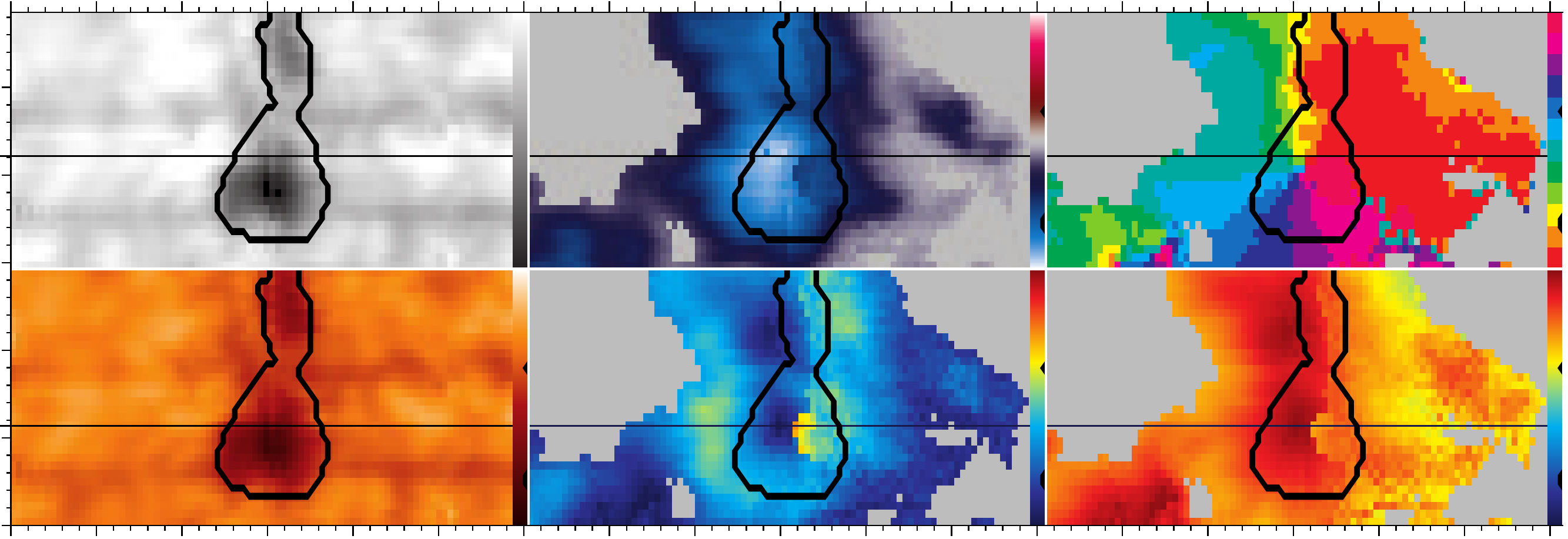}
\includegraphics[width=0.68\textwidth]{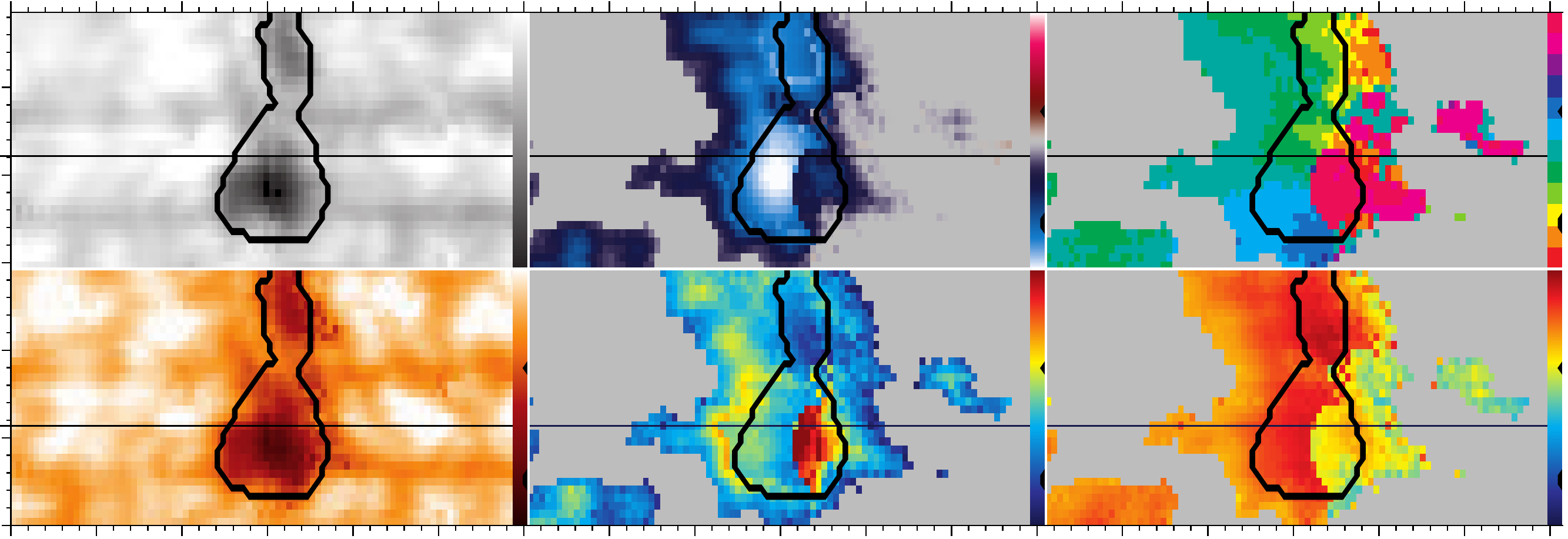}
\caption{Maps of six physical parameters derived for the Si\,\textsc{i} line 
    (\textit{top two columns}) and Ca\,\textsc{i} line (\textit{bottom two columns}) using the SIR code for the GRIS IFU mosaic starting at 07:58~UT on 6~June: normalized intensity $I / I_0$ ($0.7-1.05$), vertical component of magnetic field strength $B_z$ ($\pm$1500~G), magnetic field azimuth $\phi$ ($\pm$180\degr), temperature at $\tau = 0$ (5500~K\,--\,7000~K), horizontal component of magnetic field strength $B_\mathrm{hor}$ (0~G\,--\,1200~G), and magnetic field inclination $\gamma$ (0\degr\,--\,180\degr) (\textit{top-left to bottom-right}). The ranges for all physical parameters are depicted as colored bars on the right. The polarization signal in the gray region is below the noise level. Please note that $B_z$, $\phi$, and $\gamma$ are in the local reference frame. The major tick-marks represent 2\arcsec\ on both $x$- and $y$-axis covering the FOV of $12\arcsec \times 6\arcsec$ for each map.
    Animations showing the evolution of the pore are available online for the Si\,\textsc{i} and Ca\,\textsc{i} lines.} 
\label{FIG07}
\end{figure*}
%-------------------------------------------------------------------------------

%===============================================================================
%    Results
%===============================================================================

\section{Results\label{SEC03}}

This section covers the morphological evolution of the pores on both dates using HMI and HiFI data. Horizontal proper motions and LOS velocities are discussed. Furthermore, variations in magnetic field properties in two wavelengths, i.e., Si\,\textsc{i} and Ca\,\textsc{i}, are examined. The t-SNE projection of Si\,\textsc{i} Stokes-V profiles are discussed for both dates.

%-------------------------------------------------------------------------------
%    Evolution 
%-------------------------------------------------------------------------------

\subsection{Morphology in synoptic and high-resolution images}

Based on HMI continuum images and magnetograms the long-term evolution of both pores is retrieved. The pore observed on 29~May originated in a bipolar region. The first sign of the region was in the magnetogram at 17:00~UT on 28~May, whereas in the continuum the pore started to show up at 04:50 UT on 29~May. The leading positive polarity-pore as observed with the GRIS IFU remained visible in continuum images till 08:00~UT on 30~May but disappeared a day later at 19:00~UT on 31~May as seen in HMI magnetograms. 

In contrast, the second pore belonged to a dispersed plage region. It appeared at 06:00~UT on 6~June shortly before the GREGOR observations. The pore was located on one side of a supergranular cell whose borders were traced in negative polarity. The opposite polarity accompanying the pore was not clearly seen, however, some indication of small-scale flux emergence was present.

High-resolution G-band images allowed us to zoom in on the pores. Even though GRIS scans cover a similar FOV of the zoom-in HiFI data, the spatial resolution is not sufficient to capture the minute changes seen in the intensity images. The GREGOR time sequences cover on both days about 40 and 30~min of evolution. At the start of the HiFI sequence at 07:53~UT on 29~May the pore consisted of two umbral cores separated by a thin granular light bridge which continued to evolve. The two cores merged and were joined together with a thin dark core. The two cores were separated again by one broad granular light bridge that continued to change throughout the 40~min of observations. The large core remained till the end whereas the smaller core disappeared towards the end of the observations. G-band bright points were seen around the pore.

The pore at 07:40~UT on 6~June started also with two umbral cores separated by a light bridge, which was about one granule wide. The large bottom core retained its circular shape throughout the GREGOR observations. Over a period of 30~min, the shape and structure of the second core kept on evolving. During the sequence, the two cores appeared to merge, however, this was short-lived, and both cores separated again. As seen in the HMI magnetogram this pore was located on one side of a surpergranular cell, which was also evident in G-band images.

The total area for both pores was computed by creating a binary mask of the dark cores using an intensity threshold of $I/I_0 < 0.75$. The average area of the pore on 29~May was (10.25$\pm$0.82)~arcsec$^{2}$. When the two cores of the pore merged, the area reached up to 11.66~arcsec$^{2}$. However, the pore on 6~June has an average area of (3.52$\pm$0.38)~arcsec$^{2}$, almost three times smaller than the pore on 29~May. The maximum area covered by the 6~June pore was 4.15~arcsec$^{2}$ also during the sequence when the two cores of the pore merged. The equivalent pore radii were 1.81\arcsec\ and 1.06\arcsec\ on 29~May and 6~June, respectively. The radii were derived from the pore's area by assuming that
the pore has a circular shape. Nevertheless, on both days changes in the area of the pores were not significant throughout the 40- and 30-minute sequences, respectively.

%-------------------------------------------------------------------------------
%    Velocity pattern
%-------------------------------------------------------------------------------

\subsection{Velocity pattern}

% LCT velocity
The LCT average velocity maps for both pores are included in Fig.~\ref{FIG14}. The rainbow-colored arrows indicate the magnitude and direction of the horizontal velocity. Inflows are seen around the pore on 29~May. However, these inflows are concentrated only on the side where the pore is facing the opposite polarity. The other half of the pore showed an indication of outflows. A second layer of outflows encircled half the pore. Few scattered divergence centers are seen in the vicinity. In contrast, the pore on 6~June has very low horizontal velocity in the surroundings which mainly traced G-band bright points in the FOV. However, inflows were still present at the border of the pore. Furthermore, a few divergence centers are also present. 

For both pores, radial averages for intensity, velocity, divergence, and vorticity were computed over a radial distance of 7.5\arcsec\ and are compiled in Fig.~\ref{FIG14a}. 
The minimum intensity was lower in the pore observed on 6~June with a value of 0.3. On 29~May, the quiet-Sun intensity of $I/I_{0}=1.0$ was reached at a radial distance of about 3.1\arcsec. The intensity curve showed a dip before that, related to the separation of the two cores. For the 6~June pore, the location of $I/I_{0}=1.0$ was at the radial distance of 1.2\arcsec. An intensity enhancement can be seen immediately after that suggesting that the pore is encircled by G-band bright points. For the pore on 29~May, the maximum radial average velocity was 0.44~km~s$^{-1}$, while it was lower (0.34~km~s$^{-1}$) for the 6~June pore. The average divergence was negative in the inner region of both pores indicating strong inflows. However, the divergence was much lower for the 6~June pore with values reaching $-1.24 \times 10^{-3}$~s$^{-1}$.

% LOS velocity 
The LOS velocity for all three spectral lines was computed using the Fourier phase method \citep{Schmidt1999} with the average velocity of the quiet Sun used as the frame of reference. For the He\,\textsc{i} triplet only the red component was fitted. The maps are computed for all 20 and 30 mosaic scans taken on 29~May and 6~June, respectively. As an example, the maps in Figs.~\ref{FIG04} and~\ref{FIG05} refer to 08:02~UT on 29~May and to 07:58~UT on 6~June, respectively. The Ca\,\textsc{i} and Si\,\textsc{i} maps are scaled between $\pm$3~km~s$^{-1}$, whereas the He\,\textsc{i} maps are displayed between $\pm$20~km~s$^{-1}$. Along with the LOS velocity maps, line-core intensity maps are also compiled in the above-mentioned figures.

It is difficult to construct a clear picture of the velocity structure from the GRIS LOS velocity maps. The deep photosphere on both days depicts clearly the granulation pattern. Furthermore, as one moves to the upper photosphere, the granulation pattern dissolves somewhat, and very low velocities are traced in the core of the pore. Vague signatures of the filamentary structure are present around the pore on 29~May. However, no indications of such features are found for the 6~June pore. The line-core intensity maps for Ca\,\textsc{i} and Si\,\textsc{i} revealed some brighter regions along and inside the borders of both pores. This suggests the presence of shallower line profiles caused by strong magnetic fields. Similarly, faint indications of dark filamentary structures can also be observed in the He\,\textsc{i} line-core map for 29~May but not on 6~June. However, due to poor seeing conditions on both days, it was challenging to discern the granulation pattern and filamentary structures as seen by \citet{Verma2016b} in these infrared lines. The temporal evolution of the velocity pattern and line-core intensity is thus not clearly resolved.

%-------------------------------------------------------------------------------
%    Magnetic fields
%-------------------------------------------------------------------------------

\subsection{Magnetic field properties at two different heights}

Two-dimensional maps of physical parameters computed using two photospheric lines for both pores are compiled in Figs.~\ref{FIG06} and~\ref{FIG07}. The displayed parameters for both pores include normalized intensity (0.7\,--\,1.05), vertical magnetic field strength ($\pm 1500$~G), field azimuth ($\pm 180\degr$), temperature (\mbox{5500~}K\,--\,\mbox{7000~K}), horizontal magnetic field strength (0~G\,--\,1200~G), and field inclination (0\degr\,--\,180\degr). Only the maps at 08:02~UT and 07:58~UT are shown as examples for 29~May and 6~June, respectively. Furthermore, the temporal evolution for both pores is included in online movies.

%-------------------------------------------------------------------------------
%   Table 1
%-------------------------------------------------------------------------------
\begin{table*}
\caption{Summary of the various properties of the two pores observed on 29~May and 6~June.}
\begin{tabular}{lcc}
\hline\hline
                                       &   29~May            & 6~June\rule[-6pt]{0pt}{18pt}\\
\hline
Mean photometric area                  & (10.25$\pm$0.82)~arcsec$^2$ & (3.52$\pm$0.38)~arcsec$^2$\rule[0pt]{0pt}{12pt}\\
Mean magnetic area (Si\,\textsc{i})    &  (26.15$\pm$0.49)~arcsec$^2$  & (15.41$\pm$1.05)~arcsec$^2$ \\
Mean magnetic area (Ca\,\textsc{i})    &  (19.91$\pm$0.55)~arcsec$^2$   & (11.55$\pm$0.87)~arcsec$^2$ \\
Mean Temperature (Si\,\textsc{i})      &  (5820$\pm$90)~K       & (5540$\pm$150)~K\\
Mean Temperature (Ca\,\textsc{i})      &  (6010$\pm$120)~K      & (5630$\pm$180)~K\\
Mean B$_{z}$ (Si\,\textsc{i})          &  (1040$\pm$120)~G      & ($-$1030$\pm$210)~G\\
Mean B$_\mathrm{hor}$ (Si\,\textsc{i}) &  (380$\pm$150)~G     & (400$\pm$180)~G\\
Mean B$_{z}$ (Ca\,\textsc{i})          &  (1050$\pm$200)~G    & ($-$1030$\pm$360)~G\\
Mean B$_\mathrm{hor}$ (Ca\,\textsc{i}) &  (540$\pm$160)~G     & (640$\pm$300)~G\\
Mean LOS velocities (Si\,\textsc{i})   &  ($-$0.06$\pm$0.06)~km~s$^{-1}$ & (0.05$\pm$0.09)~km~s$^{-1}$\\
Mean LOS velocities (Ca\,\textsc{i})   &  ($-$0.20$\pm$0.49)~km~s$^{-1}$ & (0.70$\pm$0.56)~km~s$^{-1}$\\
Mean LOS velocities (He\,\textsc{i})   &  ($-$0.30$\pm$0.55)~km~s$^{-1}$ & (0.26$\pm$1.37)~km~s$^{-1}$\\
Mean Horizontal velocities             &  (0.27$\pm$0.15)~km~s$^{-1}$ & (0.30$\pm$0.13)~km~s$^{-1}$\\
Number of t-SNE clusters               & 5                    & 5\rule[-5pt]{0pt}{10pt}\\ 
\hline
\end{tabular}
\label{TAB01}
\end{table*}
%-------------------------------------------------------------------------------

The pore on 29~May has a positive vertical field strength as seen in the Si\,\textsc{i} line. The pore is surrounded by small patches of the same polarity. The lower left of the pore showed an extension towards the opposite polarity which is not obvious in the intensity images. The pore ($I/I_0 \leq 0.9$) has a mean value of $B_z= (1040 \pm 120)~G$. The horizontal magnetic field is strongest along the border of the pore. The inclination is mainly zero in the pore. The field inclination and azimuth support the field structure as seen in the horizontal and vertical magnetic field strength. The overall trend appears similar in the Si\,\textsc{i} and Ca\,\textsc{i} lines. In the Ca\,\textsc{i} line, $B_z$ has a smaller areal extent, and a strong horizontal magnetic field encircled the pore border. Overall the vertical magnetic field strength appeared similar. However, the horizontal field component in the pore has higher values in the Ca\,\textsc{i} line than in the Si\,\textsc{i} line. Apart from magnetic field properties, the temperature derived for both lines differs, i.e., the mean temperature is higher for the Ca\,\textsc{i} line. For the pores on 29~May and 6~June, the magnetic area was computed using thresholds of 700~G and $-$700~G, respectively. The mean and standard deviation for other physical parameters were computed using the magnetic area as a mask. Thus, they represent the variation of the physical properties in this area rather than a formal error estimate. Errors for inversion of Stokes profiles are difficult to quantify and no stringent and convincing work on this topic has been published to our knowledge. Different kinds of noise affect the individual Stokes profiles and optical aberrations and seeing will impact the quality of the data \citep[e.g.,][]{Lindner2020}. Furthermore, the setup of the inversions, e.g. height dependence of parameters, will affect the outcome of the inversions. If only single spectra are inverted, studying the parameter space would be advisable. However, when creating maps of physical parameters is the goal, simplicity and robustness are more important for setting up the inversions. All physical parameters describing the flow and magnetic field properties for both pores are summarized in Table~\ref{TAB01}.

The pore observed on 6~June is smaller than the one observed on 29~May. This difference is also evident in the magnetic field maps. In addition, the pore exhibits negative polarity, as indicated by the vertical field strength $B_z$. The trend observed in the physical parameters for the pore in both Si\,\textsc{i} and Ca\,\textsc{i} lines is similar to that observed on 29~May. Furthermore, the vertical and horizontal magnetic field maps reveal the extent of the field within the pore. On 6~June, strong horizontal magnetic fields are again observed at the border of the pore, with higher values than in Ca\,\textsc{i}.

\begin{figure*}[t]
\centering
\includegraphics[width=\textwidth]{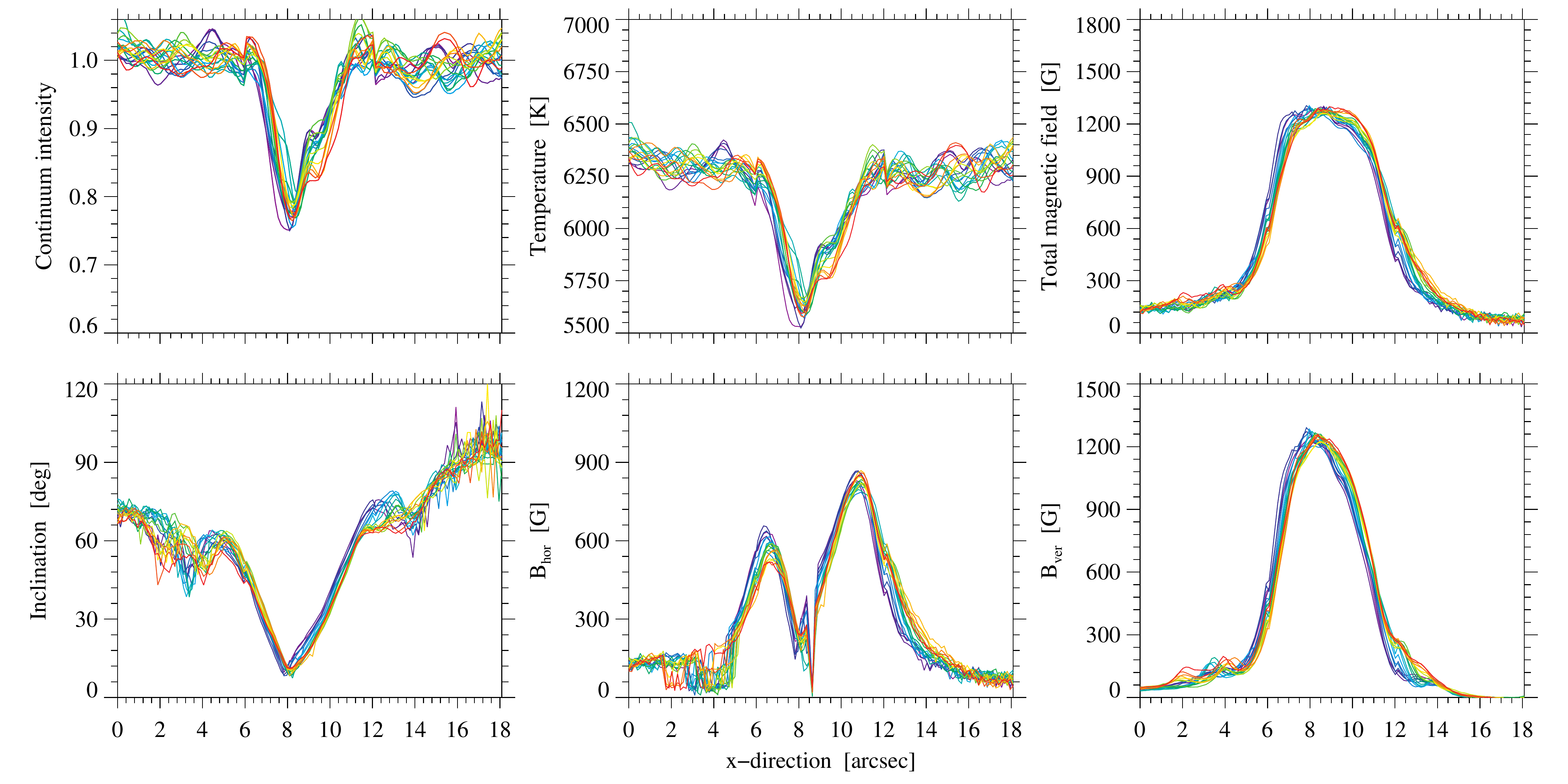}
\caption{Physical parameters estimated in the Si\,\textsc{i} line: intensity, 
    temperature, total magnetic field strength, inclination, horizontal magnetic field, and vertical magnetic field (\textit{top-left to bottom-right}) for 20 scans on 29~May  along the cutout in $x$-direction as marked in Fig.~\ref{FIG06}. The rainbow colors indicate the 20 scans with increasing order in time from blue to red.} 
\label{FIG08}
\end{figure*}

\begin{figure*}[t]
\centering
\includegraphics[width=\textwidth]{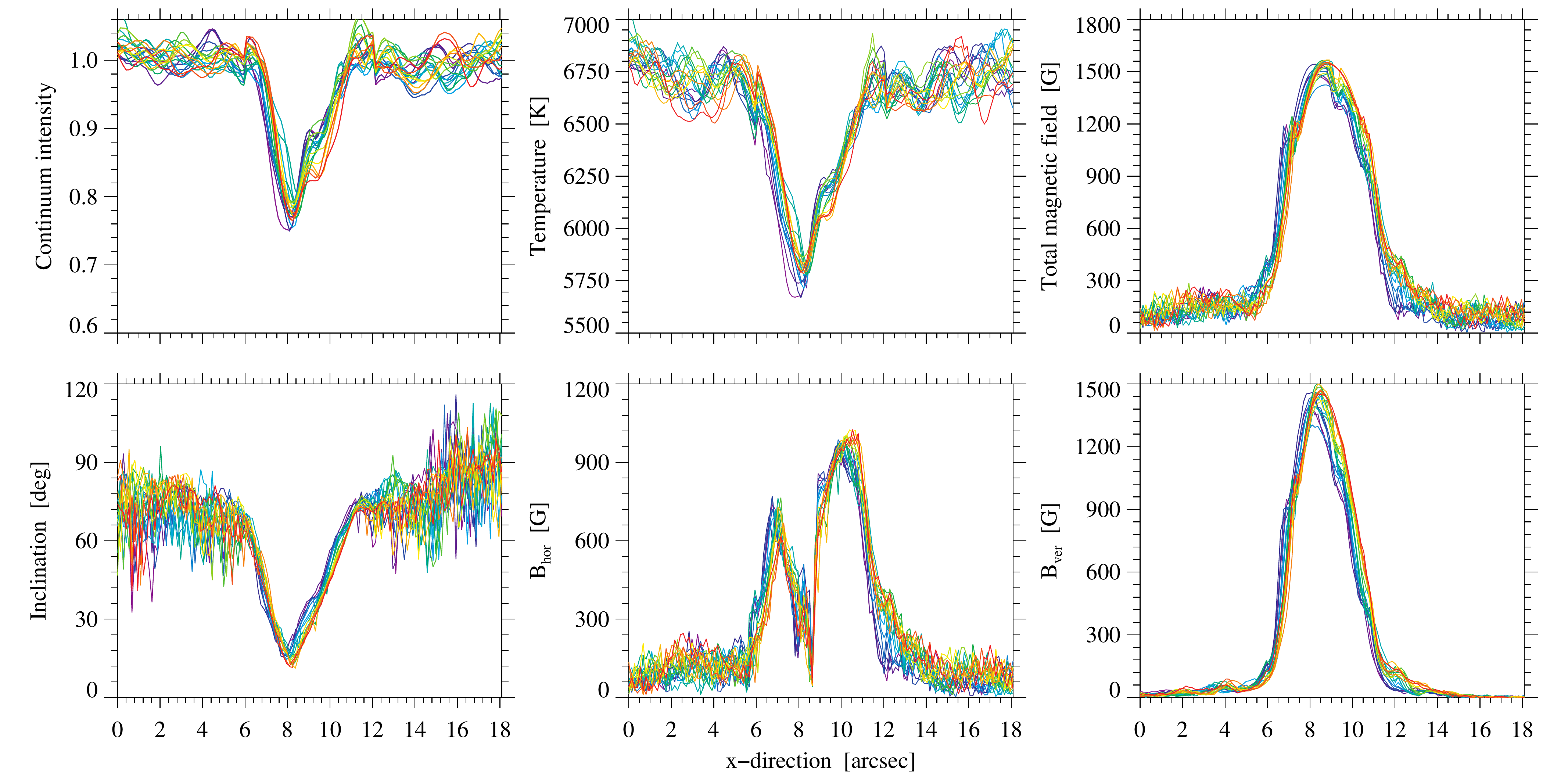}
\caption{Physical parameters estimated in the Ca\,\textsc{i} line: intensity, 
    temperature, total magnetic field strength, inclination, horizontal magnetic field, and vertical magnetic field (\textit{top-left to bottom-right}) for 20 scans on 29~May along the cutout in $x$-direction as marked in Fig.~\ref{FIG06}. The rainbow colors indicate the 20 scans with increasing order in time from blue to red.} 
\label{FIG09}
\end{figure*}

\begin{figure*}[t]
\centering
\includegraphics[width=\textwidth]{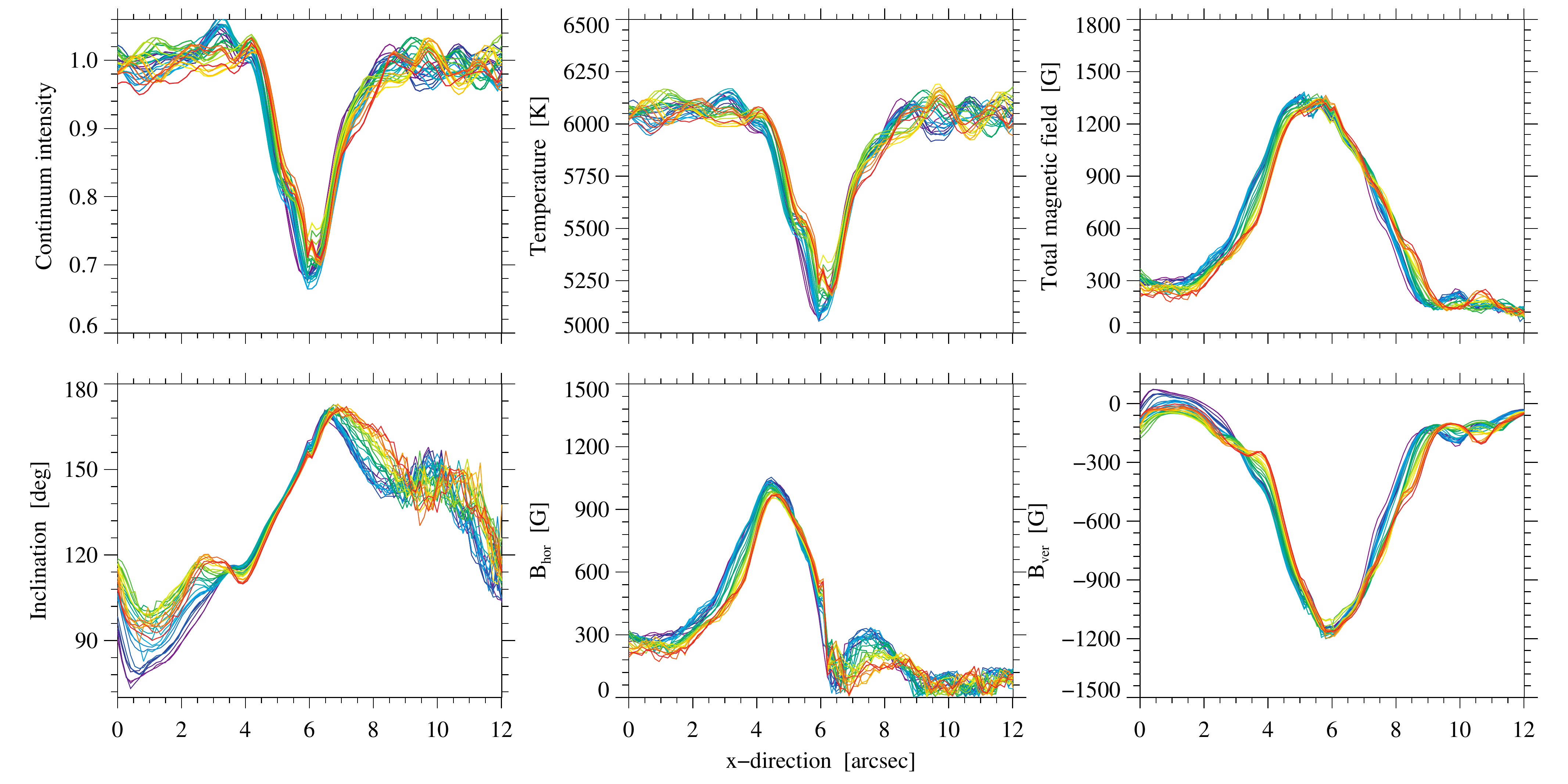}
\caption{Physical parameters estimated in the Si\,\textsc{i} line: intensity, 
    temperature, total magnetic field strength, inclination, horizontal magnetic field, and vertical magnetic field (\textit{top-left to bottom-right}) for 20 scans on 6~June along the cutout in $x$-direction as marked in Fig.~\ref{FIG07}. The rainbow colors indicate the 30 scans with increasing order in time from blue to red.} 
\label{FIG10}
\end{figure*}

\begin{figure*}[t]
\centering
\includegraphics[width=\textwidth]{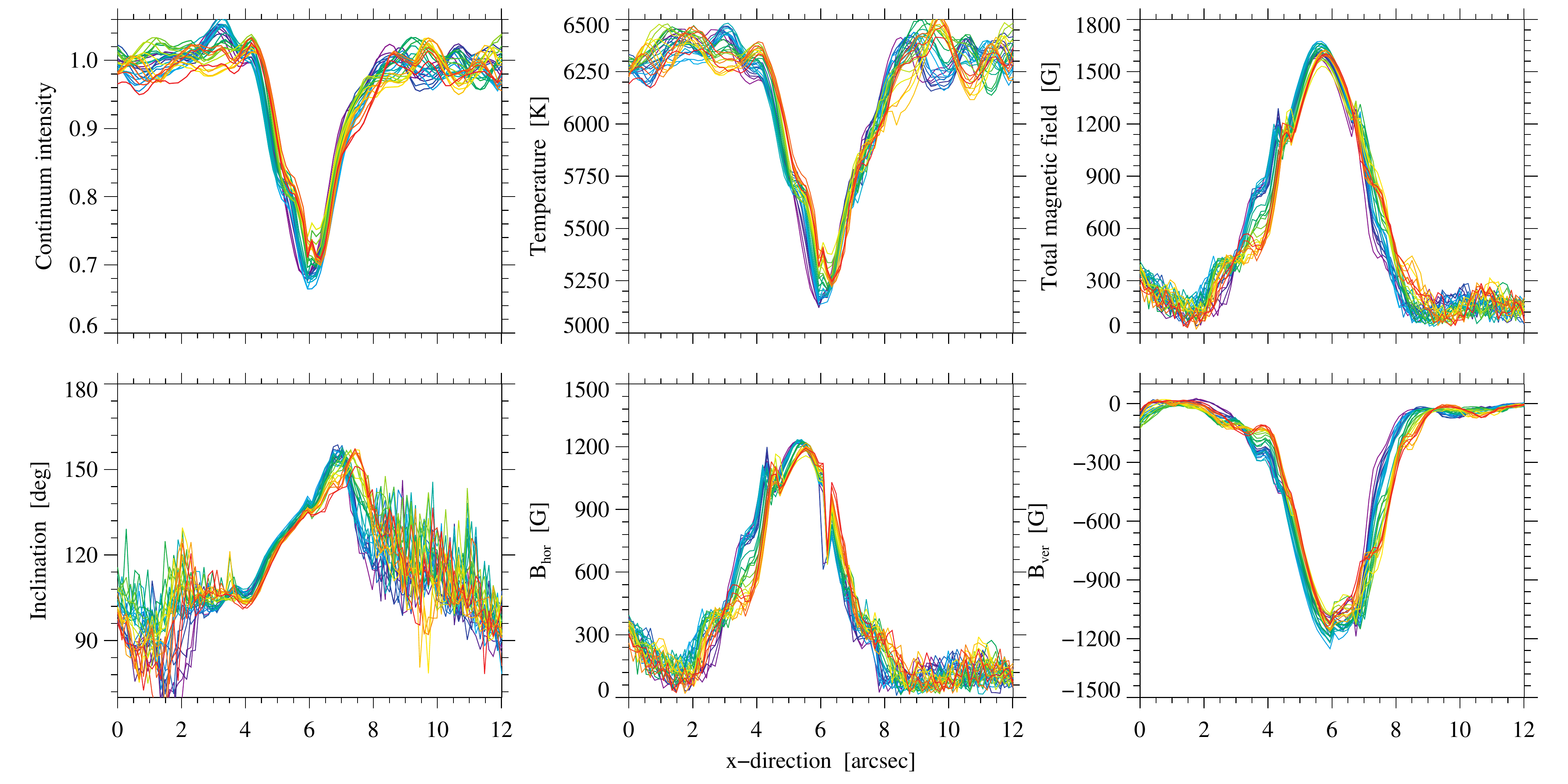}
\caption{Physical parameters estimated in the Ca\,\textsc{i} line: intensity, 
    temperature, total magnetic field strength, inclination, horizontal magnetic field, and vertical magnetic field (\textit{top-left to bottom-right}) for 30 scans on 6~June along the cutout in $x$-direction as marked in Fig.~\ref{FIG07}. The rainbow colors indicate the 30 scans with increasing order in time from blue to red.} 
\label{FIG11}
\end{figure*}

% Cutout
To visualize the temporal evolution, the variation along a cutout in the $x$-direction is displayed for both spectral lines and for both pores in Figs.~\ref{FIG08} to~\ref{FIG11}, respectively. The cutout is indicated for both pores in Figs.~\ref{FIG06} and \ref{FIG07} for 29~May and 6~June, respectively. The cutout passes through the darkest part (intensity) of the umbral core for both pores. The six displayed physical parameters are normalized intensity, temperature, total magnetic field strength, inclination, horizontal magnetic field component, and vertical magnetic field strength. The parameters are presented for all 20 and 30 mosaic scans on 29~May and 6~June, respectively, enabling a comprehensive understanding of the temporal variation of these parameters. Since the FOV of the GRIS scans is different on both days, the cutouts cover 18.2\arcsec\ and 12.2\arcsec\ on 29~May and 6~June, respectively.

%-------------------------------------------------------------------------------
%  Radial averages for 29 May
%-------------------------------------------------------------------------------
%-------------------------------------------------------------------------------

%-------------------------------------------------------------------------------
%  Radial averages for 29 May
%-------------------------------------------------------------------------------
%-------------------------------------------------------------------------------

The cutout through the normalized intensity on 29~May shows the photometric size of the pore, which is also traced in temperature. The variation of the total magnetic field appears to have the shape of a Gaussian and depicts the magnetic size of the pore. The inclination inside the pore was very low as expected but increases steadily when moving toward the border of the pore. The inclination values became very noisy in the surrounding quiet Sun, because the SIR inversions employ a mask to exclude quiet-Sun values. The horizontal component of the magnetic field has extended reach. It appears to have two maxima. However, this can result from the difficulty in estimating the horizontal component accurately in the center of the pore. The variation of the vertical component of the magnetic field appears similar to the total magnetic field. However, its FWHM is narrower than the total magnetic field. These trends are very similar for the Ca\,\textsc{i} line. Differences are in the magnetic field values and temperature. In addition, the quiet-Sun values are noisier. Furthermore, the extent of the magnetic field is narrower in the total, horizontal, and vertical components. The temporal evolution that is traced in the 20 mosaic scans is represented in the form of rainbow-colored curves where time is progressing from blue to red. No significant changes are observed for all the parameters over the period of GRIS observations. The overall trend remained the same for all parameters. However, the curves for the last few scans shown in dark blue to violet color are shifted to the left and appeared to have a somewhat lower temperature and higher values in the vertical, horizontal, and total magnetic field. This could be an indication of the further evolution of the pore.

%-------------------------------------------------------------------------------
%  Radial averages for 06 June
%-------------------------------------------------------------------------------
%-------------------------------------------------------------------------------

%-------------------------------------------------------------------------------
%  Radial averages for 06 June
%-------------------------------------------------------------------------------
%-------------------------------------------------------------------------------

For the pore on 6~June, the trend in intensity and temperature in the Si\,\textsc{i} and Ca\,\textsc{i} lines is similar to the pore on 29~May with low values inside the pore and reaching the intensity and temperature values of the quiet Sun in the surroundings. The total magnetic field also peaks at the cutout center and slowly drops down to lower values around it. The magnetic inclination is mostly above $150^\circ$ inside the pore as expected from a negative-polarity pore. In the neighboring pixels around the pore, the values are lower and noisier. The horizontal component of the magnetic field demonstrates that one side of the darkest part of the pore has higher horizontal field values than the other side. The vertical component of the magnetic field is negative for the pore with values reaching $-1200$~G in the core and with values slowly approaching quiet-Sun values outside the pore. The physical parameters in the Ca\,\textsc{i} line are very similar to those of the Si\,\textsc{i} line. The temperature has somewhat higher values in the quiet Sun. The lower temperature estimates for the Si\,\textsc{i} line could be related to difficulties in determining the continuum. This line is on one side close to the edge of the detector and on the other side, the He\,\textsc{i} triplet affects the continuum. Therefore, the inversion code arrives at a lower temperature estimate because of the incorrectly inferred lower continuum intensity. Secondly, the term quiet Sun in this context merely describes the region surrounding the pores, which may still be affected by magnetic and flow fields related to the presence of a pore. Furthermore, the extent of the total magnetic field is narrower but with higher values. The field inclination has similar values as in the Si\,\textsc{i} line but they are noisier in the neighboring quiet Sun. The horizontal component of the magnetic field is somewhat different with higher values and an almost Gaussian profile but with a steeper gradient on one side. The profile of the vertical component showed a narrow extent and a slightly higher value. Moreover, the temporal variations of the physical parameters of 30 mosaic scans depicted in rainbow colors are not significant. For both lines, a similar trend is observed but towards the middle of the sequence (light blue curves), the intensity and temperature are lower in the pore, whereas the magnetic field values are slightly higher. This coincides with the merging of two cores within the pore.

%-------------------------------------------------------------------------------
%  T-SNE over V profiles 
%-------------------------------------------------------------------------------
\begin{figure*}[t]
\centering
\includegraphics[width=0.24\textwidth]{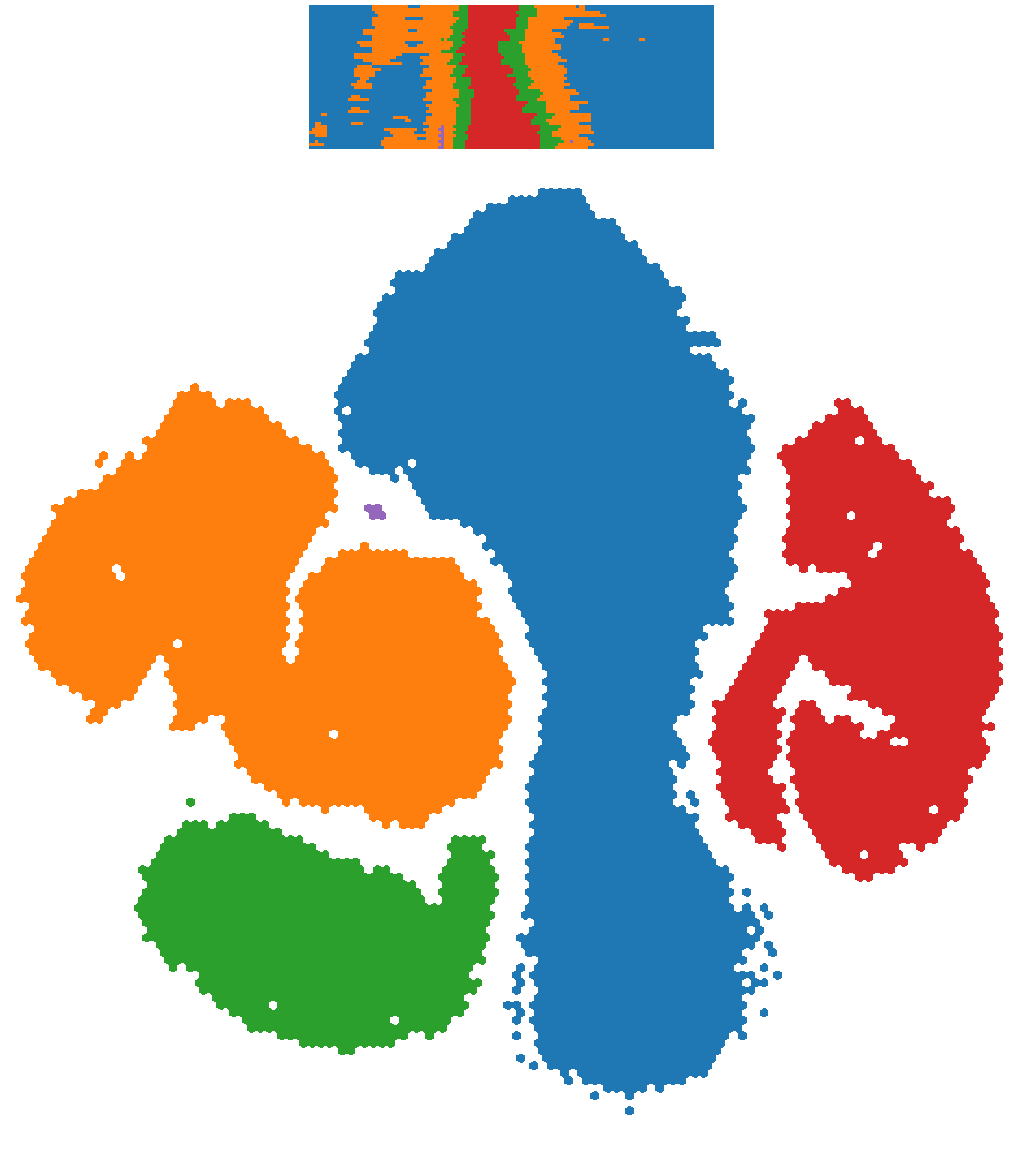}
\includegraphics[width=0.24\textwidth]{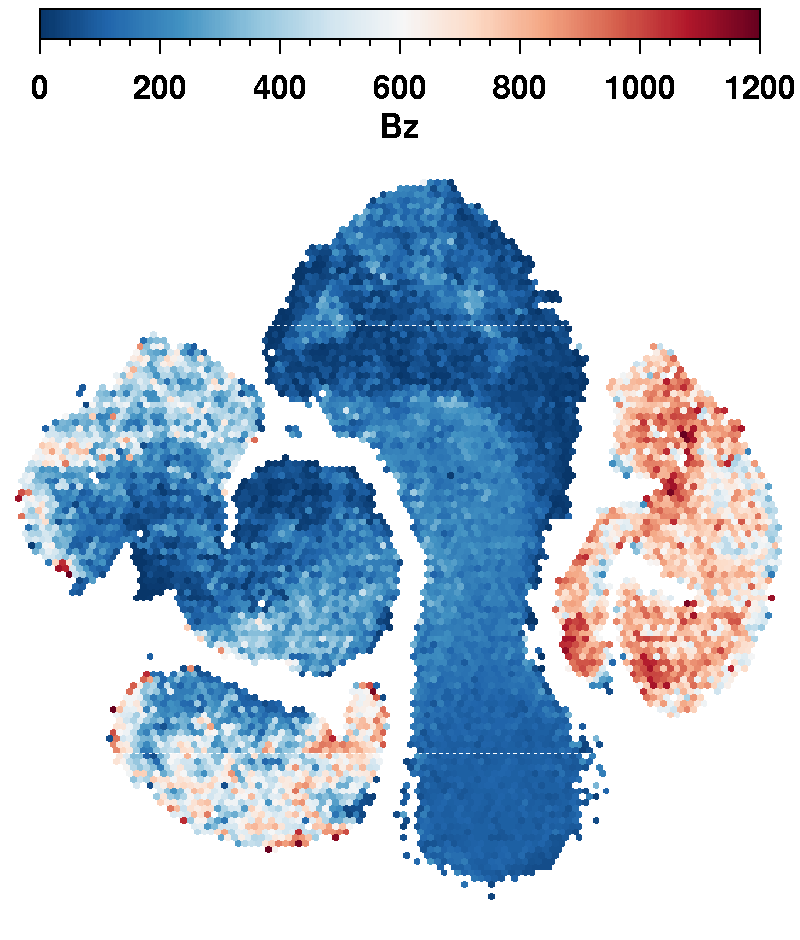}
\includegraphics[width=0.24\textwidth]{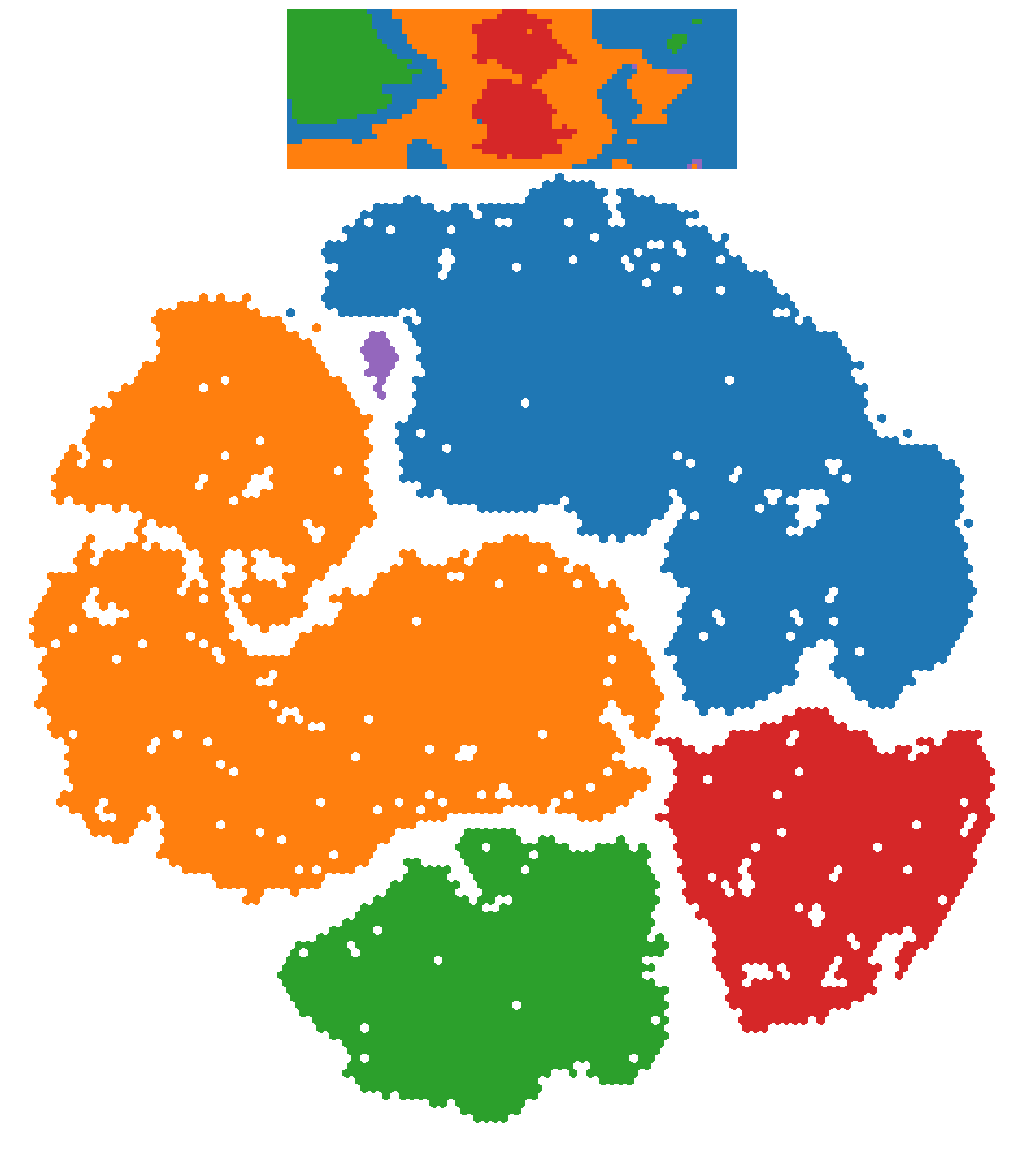}
\includegraphics[width=0.24\textwidth]{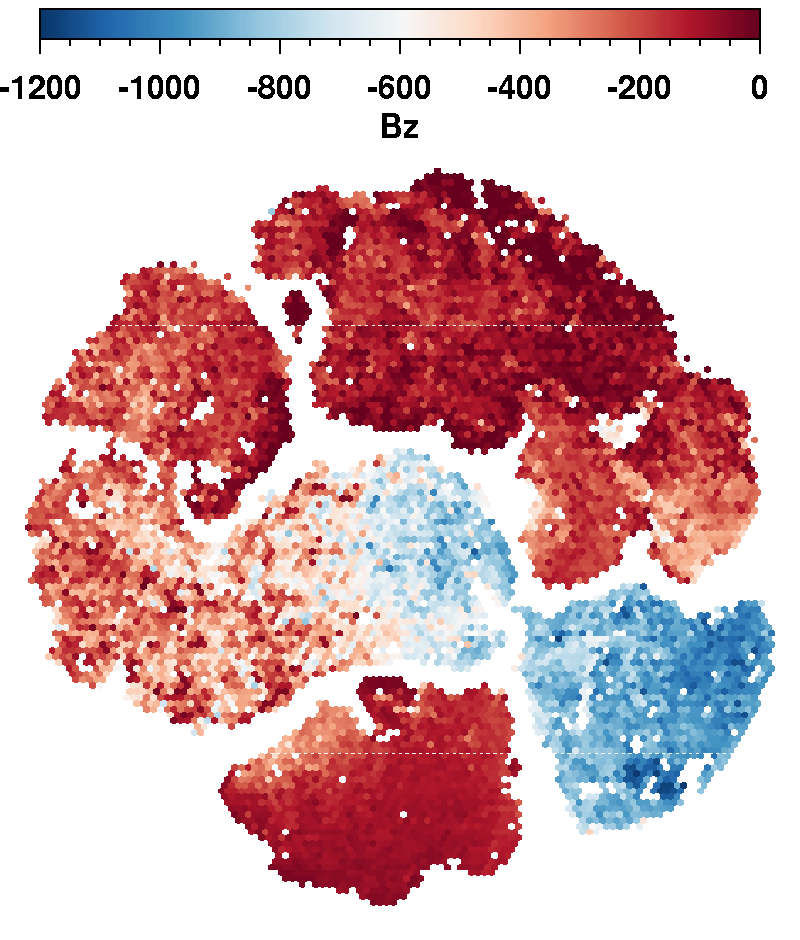}
\caption{Two-dimensional t-SNE projections based on Stokes-V profiles observed
    on 29~May (\textit{two left panels}) and on 6~June  (\textit{two right panels}). 
    The five clusters identified in the t-SNE projections for both days are color-coded.
    The vertical magnetic field strength (\textit{second and fourth panels}) is used to color the projections providing a visual guide to interpret the t-SNE projections. The back projections highlighting the onion-peel structure for both pores are shown at the top of the color-coded t-SNE projection.}
\label{FIG12}
\end{figure*}
%-------------------------------------------------------------------------------

%-------------------------------------------------------------------------------
%  TSNE profiles
%-------------------------------------------------------------------------------
\begin{figure*}
\centering
\includegraphics[width=0.98\textwidth]{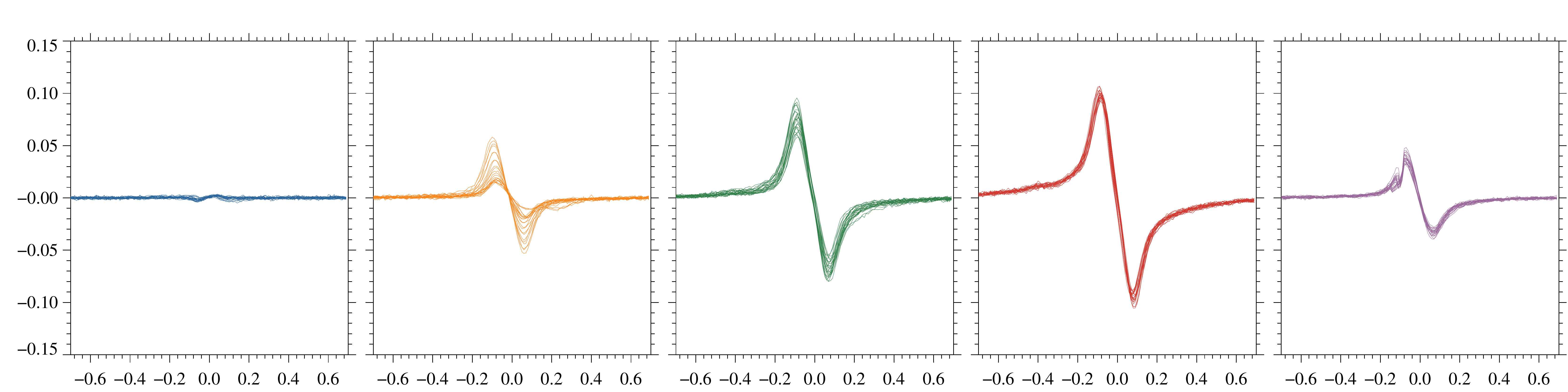}
\includegraphics[width=0.98\textwidth]{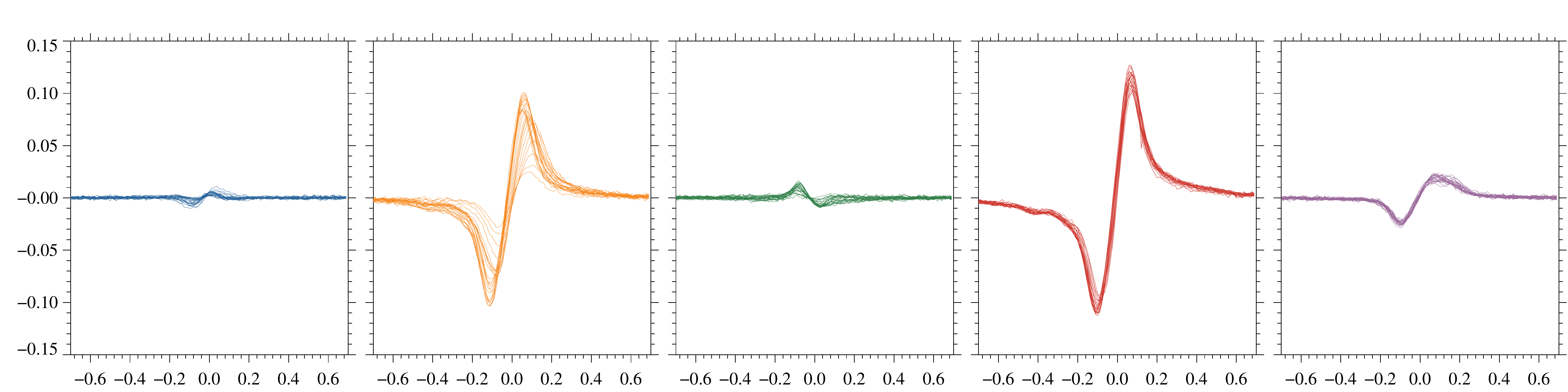}
\caption{Depicted are 25 randomly selected Stokes-V profiles for the five
    clusters of spectra in Fig.~\ref{FIG12} on 29~May (\textit{top panels}) and on 6~June (\textit{bottom panels}).} 
\label{FIG13}
\end{figure*}
%-------------------------------------------------------------------------------

%-------------------------------------------------------------------------------
%    T-SNE results on Stokes-V profiles
%-------------------------------------------------------------------------------

\subsection{Classification of spectral profiles using t-Stochastic Neighborhood Embedding?}

The two-dimensional t-SNE projections for 29~May and 6~June are compiled in Fig.~\ref{FIG12}. The Stokes-V profiles for both datasets have opposite signs. Hence, the projections are created separately to discover the clusters for each pore. The purpose of applying t-SNE was to address the question of whether the pores, which originate in different magnetic environments, have a unique distribution of Stokes profiles. Since t-SNE has been shown in previous studies to be an optimal tool for the spectral classification \citep[e.g.,][]{Verma2021}, it was also used here. Five clusters are recognized on 29~May. Out of these five, four contained the majority of the profiles, whereas the fifth cluster is the smallest. In addition, five clusters are also identified on 6~June. Twenty-five random profiles for each cluster are compiled for both days in Fig.~\ref{FIG13}. The clusters (first and third panels) in Fig.~\ref{FIG12} are marked with the color used for the profiles belonging to these clusters shown in Fig.~\ref{FIG13}.

On 29~May the profiles belonging to the largest cluster (blue, footprint shaped) are primarily associated with quiet-Sun regions (mean $I/I_{0} = 1.01 \pm 0.02$) and exhibit a very low degree of polarization. Despite their low polarization, some of these profiles display partial three-lobed structures. The profiles in the second largest cluster show a slightly higher degree of polarization and originate from the region forming an annular border between surrounding granulation and magnetic pores. The profiles in the third and fourth clusters correspond to features tracing strong magnetic fields. However, when comparing other parameters, the profiles of the third cluster (green) are mainly associated with a region having a slightly lower intensity than the quiet Sun ($0.95 \pm 0.06$) and weaker magnetic fields of $B_z = (530 \pm 380)$~G, forming a thin band around the dark core of the pore. The profiles (red) in the fourth cluster predominantly originate from the dark core of the pore, exhibiting a low mean intensity ($0.92\pm 0.06 $) and strong magnetic fields of $B_z = (740 \pm 340)$~G. The fifth and smallest cluster comprises profiles with a small hump in the blue lobe of the Stokes-V profiles. Although the other profile properties of this cluster are more similar to the second cluster, these profiles are mainly located at the border of the region belonging to the second and third clusters. The magnetic field inclination drops from $60\degr \pm 18\degr$ to $35\degr \pm 16\degr$ from the first to the fourth cluster, respectively. In summary, in the two-dimensional spatial back projection map, the first three clusters envelop the fourth cluster, akin to layers of an onion peel. The onion peel structure can be clearly seen in the insert shown in Fig.~\ref{FIG12}. The displayed back-projected map corresponds to the scan taken at 08:02~UT.

% 6 June 2019
The 6~June profiles also exhibit distinct appearances and properties forming five clusters. Similar to the profiles on 29~May, the largest cluster consists of profiles associated with the quiet-Sun intensity (mean $I/I_{0} = 1.00 \pm 0.03$) and weak magnetic fields of $B_z = (-130 \pm 180)$~G. Profiles in the second largest cluster display a two-lobed structure with moderately strong magnetic fields of $B_z = (-380 \pm 320)$~G, transitioning from the mean quiet-Sun intensity to slightly lower intensities ($I/I_{0} = 0.98 \pm 0.06$). These profiles primarily originate from the band encircling the dark core, separating it from the surrounding quiet Sun. The third largest cluster comprises three-lobed profiles with opposite-polarity features of very weak magnetic strength, mainly originating from quiet-Sun regions that face solar disk center. Profiles in the fourth cluster are well-separated, two-lobed, and primarily belong to the dark core ($I/I_{0} = 0.88 \pm 0.08$) of the pore, which is characterized by very strong magnetic fields of $B_z = (-900 \pm 180)$~G. The fifth and smallest cluster consists of two-lobed profiles with a clear slanted connection between the two lobes. These profiles originate at the border of the region belonging to the first and second clusters. The magnetic field inclination increases from $126\degr \pm 20\degr$ for the first cluster to $133\degr \pm 17\degr$ for the second cluster which further goes up to $159\degr \pm 10$ for the fourth cluster. The third cluster has a similar inclination as the second cluster. However, the inclination is lowest for the fifth cluster with values of $118\degr \pm 9\degr$. Similar to the pore observed on 29~May, the pore on 6~June also exhibits an onion peel structure, where regions belonging to the first and second clusters encircle the region of the fourth cluster as seen in the sample back projected map at 07:58~UT in Fig.~\ref{FIG12}. Note that the color coding and distribution of the clusters are not identical for both pores. The 6~June profiles trace only two clusters around the pore, while the 29~May profiles trace three clusters around the pore.

%===============================================================================
%    Discussion
%===============================================================================

\section{Discussion and conclusions\label{SEC04}}

This study presents the evolution of two pores using high-resolution GREGOR observations which are complemented by HMI continuum images and magnetograms. The first pore, observed on 29~May, originated in a bipolar region and was visible until 30~May. The second pore, observed on 6~June, belonged to a dispersed plage region and appeared shortly before the GREGOR observations. None of the pores developed a penumbra during their entire lifetime. In HiFI images G-band bright points were observed around both pores. High-resolution G-band images revealed minute changes in intensity which were not captured by GRIS scans. The 29~May pore consisted of two umbral cores separated by a thin granular light bridge, which evolved and the cores merged during the observations. The 6~June pore also had two umbral cores with a connecting dark lane where the pores also merged but then separated during observations. The pore on 6~June has an area almost three times smaller than the 29~May pore. During the 40-minute sequence on 29~May and the 30-minute sequence on 6~June, the changes in both the area and other physical parameters of the pores were found to be insignificant. 

Both pores can be categorized as smaller pores as mentioned in the statistical study by \citet{Verma2014}, where 75\%/66\% (single/time-averaged G-band images) of all pores were smaller than 5~Mm$^2$, which is comparable to an area of just a few granules. \citet{Verma2014} concluded that smaller pores tend to be more circular. However, both observed pores were not circular and both had corrugated boundaries. The high-resolution G-band images also reveal thin hair-like structures \citep{Scharmer2002} around both observed pores.

The G-band LCT maps show persistent inflows around both pores. Inflows based on horizontal plasma velocities were previously observed \citep[e.g.,][]{Kamlah2023, Sobotka2012, Hirzberger2003, Roudier2002} and also reported in simulations \citep[e.g.,][]{Cameron2007}. However, the values differ in all studies. This can be related to different techniques to measure optical flows and the respective parameter choices for a given technique. However, the horizontal velocities in close proximity to both pores are still on the lower side compared to previous studies. On 6~June, the pore exhibited significantly lower velocities than the 29~May pore, which can be attributed to the dispersed magnetic field surrounding the pore, as signified by the adjacent G-band bright points. Although the G-band bright points are seen around both pores, their enhanced presence around the pore on 6~June is traced as an intensity enhancement in the radial intensity average (Fig.~\ref{FIG14a}). The observed inflows are related to downflows at the borders of both the pores as seen in the LOS velocity maps for the Si\,\textsc{i} and Ca\,\textsc{i} lines, which agrees with previous studies \citep[e.g.,][]{Chae2015, Roudier2002}.

The various physical parameters inside these two pores were retrieved using SIR inversion. \citet{Quintero2016} utilized deconvolution and inversion technique to derive the physical properties of a pore. Our results align with their findings, confirming that the temperature inside both pores was lower than the surrounding area, and the magnetic field strength exceeded 1500\,--\,1800~G. In addition, the magnetic field strength decreases with height. On average the field strength values are lower for the Si\,\textsc{i} line than for the Ca\,\textsc{i} line, where the latter originates deeper in the photosphere, whereas the spatial extent of the magnetic structures increases for the Si\,\textsc{i} line. The chromospheric properties within and around the pores play a crucial role in the transition of a pore to a sunspot \citep[e.g.,][]{Kamlah2023, Sobotka2013}. However, this aspect was not extensively investigated as none of the observed pores developed a penumbra. The LOS velocities and line-core intensity computed for chromospheric He\,\textsc{i} triplet do exhibit some filamentary structure for the large pore on 29~May. Conversely, such a pattern is absent in the maps for the pore observed on 6~June.

The application of t-SNE facilitates identifying unique clusters within the spectral data, as demonstrated in previous studies \citep[e.g.,][]{Verma2021, Matijevic2017}. Analysis of the Stokes-V profiles for the 29~May and 6~June observations revealed the presence of five distinct profile classes within each pore. Upon back-projection, the resulting structure of the magnetic field exhibited an onion peel arrangement, with clusters of increasing magnetic field strengths surrounding one another. Notably, certain profiles exhibited a unique shape and originated at the border of the quiet Sun and a region with high magnetic field strength. In both pores, a particular cluster (first for 29~May and third for 6~June) exhibited three-lobed Stokes-V profiles. These three-lobed profiles suggested the presence of plasma characterized by the simultaneous existence of very weak magnetic fields with opposite polarities, accompanied by flows directed in opposite directions \citep{Kiess2018}. However, only the pore on 6~June has a cluster (green colored) with very weak Stokes-V profiles of opposite sign than the main pore. This could be an indication that despite appearing to be isolated the surrounding of that pore contained opposite polarity magnetic fields. Remarkably, despite differences in size, polarity, and origin environment, t-SNE yielded a similar classification for both pores. 

Most of the observational results such as low intensity and vertical magnetic fields inside pores are in agreement with the numerical modeling results \citep{Stein2011, Cameron2007}. \citet{Cameron2007} found that the smaller pores are brighter. However, the pore on 6 June, which is smaller, has a much lower intensity as seen in both imaging and spectropolarimetric data. Although the intensity values at different wavelengths are of the same order as mentioned in \citet{Cameron2007}. The low intensity could be related to the high filling factor in the smaller pore \citep{Sobotka1999}. The field inclination (with respect to the surface normal) in and around both pores reduces at their borders. On average the inclination changes by about 30\degr\,--\,35\degr\ in the case of both pores from the center to the edges of the pores. This trend is also in agreement with simulations and past observations \citep[e.g.,][]{Cameron2007, Suetterlin1998}.

The following list summarizes the results with respect to the working hypothesis that different mechanisms (e.g., local vs.\ global dynamo action) in different environments create isolated and active-region pores.

\begin{itemize}
\item Both pores were separated by a thin granular light bridge during their evolution, and the borders of the pores were corrugated, indicating initial interactions of the magnetic field with convective plasma motions. However, these interactions remained weak, and the pores never developed a penumbra.
\item The photometric areas of 3.5 arcsec$^2$ and 10.3 arcsec$^2$ for the isolated and active-region pores, respectively, are on the smaller side with respect to the size distribution given in \citet{Verma2014}. The core intensity of the isolated pore was lower despite its smaller size.
\item Inflows around both pores are present in average horizontal flow maps. Radial averages of intensity, velocity, divergence, and vorticity indicate stronger inflows for the active-region pore.
\item Both pores show the surrounding granulation pattern and low velocities in the photospheric LOS velocity maps. Signs of filamentary structures are present in the chromospheric LOS maps of the active-region pore. This raises the question whether signatures of penumbral fields first become visible in the photosphere or in the chromosphere.
\item The magnetic fields of both pores had comparable mean values and were fairly stable during the observing period, with only small variations indicating evolutionary changes. Both pores have strong vertical internal magnetic fields and weaker, more inclined fields along their borders.
\item The magnetic field properties are similar in the Si\,\textsc{i} and Ca\,\textsc{i} lines. The pores have higher magnetic field values and smaller magnetic areas in the Ca\,\textsc{i} line compared to the Si\,\textsc{i} line, reflecting the canopy effect for expanding flux tubes.
\item The two-dimensional t-SNE projections of the Si\,\textsc{i} Stokes-V profiles show five clusters. Their back projection reveals an onion-peel structure for both pores, where clustered Stokes-V profiles indicate concentric layers with different magnetic field properties. 
\item The outermost layer belongs to the surrounding quiet Sun, and the innermost layer represents the core of the pores, which is surrounded by another layer with weaker magnetic field values than the core but higher values than the quiet Sun.
\item The active-region pore has an additional layer with slightly lower intensity than the quiet Sun and weaker magnetic fields than the inner core of the pore, indicating a subtle difference between the magnetic field structure of the active-region and isolated pores.
\item The power of t-SNE as a diagnostic tool was demonstrated in this case study. Even in this small dataset, manual/visual discrimination between different classes of Stokes profiles becomes impossible. Only larger, more diverse, and statistically meaningful samples will reveal the full potential of t-SNE.
\end{itemize}

The transition from a pore to a sunspot, i.e., the formation of a penumbra, still requires further investigation. The observed isolated and active-region pores shared many similarities but the active-region pore already showed some features favoring penumbra formation. However, case studies are necessarily affected by selection effects, which in this case did not allow us to validate the working hypothesis. Only the pore size may be sufficient to explain the observed discrepancies. Nevertheless, the current results encourage further research by extending the GRIS IFU observations to large and more complex active-region pores and to less complex isolated pores (i.e., no granular or faint light bridges), covering a wider size range including some of the largest isolated pores. Validation of the working hypothesis may thus still be possible.

The idea of a fast imager in combination with an IFU is to capture the evolution of solar features at a high cadence and instantaneously with high-spatial and high-spectral resolution, respectively. The high-resolution HiFI images approach the diffraction limit of the 1.5-meter GREGOR solar telescope, and the observed images closely resemble those of simulated pores in the work of \citet{Cameron2007}. The GRIS IFU shows promising results at the GREGOR solar telescope. It demonstrated its ability to provide near-infrared spectropolarimetry of rapidly evolving solar magnetic features. The IFU mosaics with a FOV of $18\arcsec \times 9\arcsec$ and $12\arcsec \times 6\arcsec$ for both pores were acquired on average in two and one minute(s) on 29~May and 6~June, respectively, which is faster than a traditional slit spectropolarimeter \citep{DominguezTagle2022}. However, in the case of the observed pores, the variation of the physical parameters due to the temporal evolution was not significant. Furthermore, the seeing was variable during the IFU mosaics on both days, which resulted in a lower quality of the data. This together with the relatively short observing period with the GRIS/IFU dataset posed some limits on this investigation of pores. Under excellent seeing conditions, observations of up to two hours are possible. The same argument applies to observations at the best mountain-island observing sites. However, to have a comprehensive understanding of pore formation and evolution along with precisely measured physical properties, more and better data are needed. With the 4-meter Daniel K.\ Inouye Solar Telescope \citep[DKIST,][]{Rimmele2020}, tracking features in multiple spectral lines becomes possible with an extraordinary spatial resolution,  which promises to reveal details that are currently only accessible in numerical simulations \citep{Rempel2014}. 
Moreover, the first results of the GRIS IFU support the choice of this type of instrument for three-dimensional spectropolarimetry in the proposed 4-meter European Solar Telescope \citep[EST,][]{Quintero2022}.

%===============================================================================
%    Acknowledgements
%===============================================================================

\begin{acknowledgements}
The 1.5-meter GREGOR solar telescope was built by a German consortium under the leadership of the Institute for Solar Physics (KIS) in Freiburg with the Leibniz Institute for Astrophysics Potsdam (AIP), the Institute for Astrophysics G\"ottingen, and the Max Planck Institute for Solar System Research (MPS) in G\"ottingen as partners, and with contributions by the Instituto de Astrof\'{\i}sica de Canarias (IAC) and the Astronomical Institute of the Academy of Sciences of the Czech Republic (ASU). HMI data are provided by the Joint Science Operations Center -- Science Data Processing. This study was supported by the European Commission’s Horizon 2020 Program under grant agreement 824135 (SOLARNET).
MV acknowledges support by grant VE~1112/1-1 of the Deutsche Forschungsgemeinschaft (DFG). 
MV would like to thank the anonymous referee for carefully reading the manuscript and providing ideas, which significantly enhanced the paper.
\end{acknowledgements}

%===============================================================================
%    Bibliography
%===============================================================================

\end{document}